\documentclass[twoside,11pt]{article}
\usepackage{jair, theapa, rawfonts}
\usepackage{graphicx}
\usepackage{comment}

\usepackage[T1]{fontenc}
\usepackage{tikz}
\usetikzlibrary{arrows, positioning, automata,shapes}

\usepackage{float}
\usepackage{graphicx}
\usepackage{amssymb,amsmath,amsfonts, amsthm}
\allowdisplaybreaks[1]
\usepackage{nicefrac}       
\usepackage{microtype}      
\usepackage{tcolorbox}
\usepackage{xcolor}
\usepackage{empheq}     

\usepackage[moderate]{savetrees}

\newtheorem{theorem}{Theorem}

\newtheorem{lemma}{Lemma}

\newtheorem{fact}{Fact}

\newtheorem{assumption}{Assumption}

\newcommand{\Gr}{{\cal V}^{\sigma}}
\newcommand{\Hy}{{\cal V}^{\psi}}
\newcommand{\Op}{{\cal V}^*}
\newcommand{\B}{{\cal B}}
\newcommand{\A}{{\cal A}}

\newcommand{\boxedeq}[2]{\begin{empheq}[box={\fboxsep=6pt\fbox}]{align}\label{#1}#2\end{empheq}}

\newcommand{\E}{\mathbf{E}}
\DeclarePairedDelimiter{\cardinality}{\vert}{\vert}

\jairheading{71}{2021}{347-370}{08/2020}{06/2021}
\ShortHeadings{Stochastic Submodular Cover}
{Hellerstein, Kletenik, \& Parthasarathy}
\firstpageno{347}

\begin{document}

\title{A Tight Bound for Stochastic Submodular Cover}

\author{\name Lisa Hellerstein \email lisa.hellerstein@nyu.edu \\
\addr Dept. of Computer Science and Engineering \\
NYU Tandon School of Engineering \\
Brooklyn, NY 11201 USA
       \AND
       \name Devorah Kletenik \email kletenik@sci.brooklyn.cuny.edu \\
     \addr Dept. of Computer and Information Science \\
     Brooklyn College, City University of New York \\
     Brooklyn, NY 11210 USA
       \AND
       \name Srinivasan Parthasarathy \email spartha@us.ibm.com \\
       \addr IBM T. J. Watson Research Center \\
       Yorktown Heights, NY 10598 USA}


\maketitle

\begin{abstract} 
We show that the Adaptive Greedy algorithm of Golovin and Krause achieves an approximation bound of $(\ln (Q/\eta)+1)$
for Stochastic Submodular Cover: here $Q$ is the ``goal value'' and $\eta$ is the minimum gap between $Q$ and any attainable utility value $Q' < Q$.
Although this bound was claimed by Golovin and Krause in the original version of their paper, the proof was later shown to be incorrect by Nan and Saligrama.  The subsequent corrected proof of Golovin and Krause gives a quadratic bound of $(\ln(Q/\eta) + 1)^2$.   
A bound of
$56(\ln(Q/\eta) + 1)$ is implied by work of Im et al. 
Other bounds for the problem depend on quantities other than $Q$ and $\eta$. 
Our bound restores the original bound claimed by Golovin and Krause, generalizing the well-known $(\ln~m + 1)$ approximation bound on the greedy algorithm for the classical Set Cover problem, where $m$ is the size of the ground set.
\end{abstract}

\section{Introduction}
\label{Introduction}

The {\em Stochastic Submodular Cover} problem was introduced by Golovin and Krause~\citeyear{Golovin:2011:AST:2208436.2208448}.  Their interest in this problem was motivated by applications in a variety of areas, including sensor placement, viral marketing, and active learning.
The problem combines and generalizes two previously studied generalizations of the classical NP-complete Set Cover problem: {\em Submodular Cover} and {\em Stochastic Set Cover}.

\begin{figure}[hbt]
\centering
\begin{tikzpicture}[>=stealth',shorten >=5pt,semithick, node distance=3cm,on grid,initial/.style={}
]
\tikzset{edge/.style = {->,> = latex'}}
 \node (CSC) {Classical Set Cover};
    \node (SbSC) [above right =3.5cm of CSC] {Submodular Cover};
    \node (StSC) [above left  =3.5cm of CSC] {Stochastic Set Cover};
    \node (SSSC) [above right =3.5 of StSC] {Stochastic Submodular Cover};
    
    \path[->] (CSC) edge  (SbSC);
    \path[->] (CSC) edge (StSC);
    \path[->] (StSC) edge  (SSSC);
    \path[->] (SbSC) edge (SSSC);
\end{tikzpicture}
\caption{Relationship between cover problems}
\label{fig:set-cover}

\end{figure}
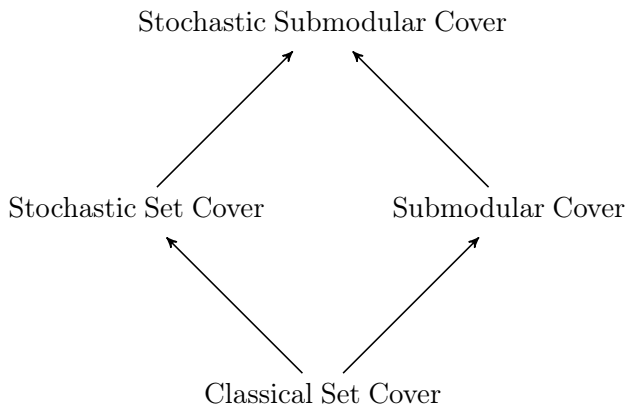

In the Submodular Cover problem, introduced by Wolsey~\citeyear{wolsey:submod}, the input
consists of a finite set of {\em items} $E$, a utility function $f:2^{E} \rightarrow \mathbb{R}_{\geq 0}$
(given by an oracle), and a cost function $c:E \rightarrow \mathbb{R}_{>0}$.  The function $f$ 
is a {\em polymatroid function}, that is, $f(\emptyset) = 0$ and $f$ is both {\em monotone} and {\em submodular}.  Monotonicity means that
$\forall A \subseteq B \subseteq E,~  f(A) \leq f(B)$.  Submodularity means that
\begin{gather*}
  \forall A \subseteq B \subseteq E,~\forall e \in E \setminus B \nonumber \\
  f(A \cup \{e\}) - f(A) \geq f(B \cup \{e\}) - f(B) 
\end{gather*}
The goal in the Submodular Cover problem is to find a subset $A \subseteq E$ such that $f(A) =  f(E)$ and
$\sum_{e \in A} c(e)$ is minimized. 

The fact that the Submodular Cover problem is a generalization of the classical Set Cover problem can be seen as follows.  In the classical Set Cover problem, the input consists of a ground set of elements $D = \{ d_1, \ldots, d_m\}$, a family of subsets of the ground elements, $\mathcal{S} = \{S_1, \ldots, S_n\}$ whose union is equal to $D$, and a cost function $c:\mathcal{S} \rightarrow \mathbb{R}_{>0}$.  View each $S_i \in \mathcal{S}$ as an item, and define the polymatroid utility function $f:2^{\mathcal{S}} \rightarrow \mathbb{R}_{\geq 0}$ such that for $A \subseteq \mathcal{S}$, $f(A) = |\bigcup_{S_i \in A} S_i|$.  Finding a min-cost subset of items $A$ such that $f(A)= f(\mathcal{S})$ is equivalent to finding a min-cost cover of the ground elements. 

Wolsey's greedy algorithm for Submodular Set Cover repeatedly selects the item yielding the largest increase in utility, per unit cost.  Wolsey's analysis of this algorithm shows that it achieves an approximation bound of $(\ln (Q/\eta) + 1)$, where $Q = f(E)$ and $\eta$ is the minimum gap between $Q$ and any attainable utility value $Q' < Q$.

The Stochastic Set Cover problem is a different generalization of the classical Set Cover problem, which captures situations in which there is uncertainty about which ground elements will be covered by each ``subset''. 
A motivating example for this problem is a sensor placement problem, where prior to placing a sensor in a given location, it is unclear which subset of locations will be monitored by the sensor
~\cite{Golovin:2011:AST:2208436.2208448,goemans:stochasticcovering}.  
The input to the Stochastic Set Cover problem consists of a ground set $D = \{d_1, \ldots, d_m\}$, a family $\mathcal{S} = \{S_1, \ldots, S_n\}$ of stochastic subsets of the ground set, and a cost function $c:\mathcal{S} \rightarrow \mathbb{R}_{>0}$.
There is an integer $k > 0$, such that
each $S_i$ is an independent random variable, whose value can be any one of $k$ different subsets of ground set $D$ (the $k$ subsets can be different for each $i$).
For each of the $k^n$ possible combinations of 
values of the $n$ variables $S_i$, it holds that $\bigcup_{i=1}^n S_i = D$.
Choosing $S_i$ incurs cost $c(S_i)$, the value of $S_i$ is not revealed until $S_i$ is chosen, and each $S_i$ can be chosen at most once.  
The problem is to determine how to sequentially and adaptively select stochastic subsets $S_i$, until the union of the (values of the) chosen $S_i$ is equal to $D$, so as to minimize the expected cost incurred.
Selecting ``adaptively'' means that the choice of the next subset at each step can depend on the revealed values of the $S_i$'s that were already chosen (but not on the values of the other $S_i$'s).   

A natural greedy policy for this problem is to repeatedly choose the stochastic subset which 
is expected to cover the largest number of uncovered items, per unit cost.
Liu et al.~\citeyear{Liu:2008:NAS:1376616.1376633} presented and analyzed this policy for a special case of Stochastic Set Cover. 
They claimed that the expected cost incurred by this policy is at most a factor of $(\ln~m + 1)$ greater than the optimal expected cost of any other sequential, adaptive policy for the problem.  
However, as we describe in Appendix~\ref{app:error}, the analysis of Liu et al.\ was not correct.

The problem addressed in this paper, the Stochastic Submodular Cover problem, combines and generalizes the Stochastic Set Cover problem and the Submodular Cover problem.\footnote{When Wolsey introduced the Submodular Cover problem, he called it Submodular Set Cover.  Similarly, Stochastic Submodular Cover is sometimes called Stochastic Submodular Set Cover.}
We give an informal definition of the problem here, and a more formal definition in Section~\ref{sec:prelim}.
In this problem, there is a set of stochastic items $E$, 
and a cost function $c:E \rightarrow \mathbb{R}_{>0}$.
Each item can be in one of $k$ possible ``states''. The states of the different items are independent.
There is a utility function $f$ that assigns a real value to each subset $A$ of items from $E$.  This value depends not only on which items are in $A$, but also on the states of those items. The value of $f$ on the empty set is 0.  For any fixed assignment
of states to items,
$f$ is monotone and submodular.  On the set of all items in $E$, the value of $f$ is guaranteed to be equal to a fixed value $Q$, called the {\em goal value}, regardless of the states of the items.
The state of an item
$e \in E$ is not revealed until item $e$ is chosen and its cost $c(e)$ is incurred. Each item can be chosen at most once.   
The problem is to find a policy for sequentially selecting items from $E$ until the value of $f$ on the set of chosen items (in their revealed states) is equal to $Q$, so as to minimize the expected cost incurred.  
The policy can be adaptive, meaning that the choice of the next item to select can depend on the states of the items already chosen (but not on the states of the other items).

We say that an algorithm achieves a $\gamma$-approximation bound for this problem, where $\gamma \geq 1$, if it produces a sequential, adaptive policy whose expected cost is at most a factor of $\gamma$ larger than the minimum expected cost of any sequential, adaptive policy.  

Golovin and Krause~\citeyear{Golovin:2011:AST:2208436.2208448} proposed and analyzed a greedy algorithm for Stochastic Submodular Cover, called
Adaptive Greedy.  It repeatedly selects the item which would yield the largest expected increase in utility, per unit cost.  
This generalizes the greedy algorithms described above. Golovin and Krause also considered a setting where the Adaptive Greedy algorithm does not make an exactly optimal choice in each greedy step, but instead makes an $\alpha$-approximate greedy choice.
They claimed that Adaptive Greedy achieves an approximation bound of $\alpha(\ln (Q/\eta) + 1)$, but
subsequently Nan and Saligrama~\citeyear{nanSaligrama} showed that  their proof of the bound was  incorrect. (The error in Golovin and Krause's proof is not the same as the one in the paper of Liu et al., and the proofs use different approaches.)  In a later version of their paper, Golovin and Krause~\citeyear{GolovinKrauseCorrected} presented a corrected version of their proof, but with a weaker quadratic bound of $\alpha(\ln (Q/\eta) + 1)^2$.

Im et al.~\citeyear{imetal} studied a problem called Weighted Stochastic Submodular Ranking, which generalizes Stochastic Submodular Cover.  
Their work implies that Adaptive Greedy, with $\alpha=1$, achieves a 
bound of $56(\ln (Q/\eta) + 1)$ for Stochastic Submodular Cover.
The proof technique used by Im et al.\ relies on a latency argument with geometrically increasing time intervals. 
They did not attempt to minimize the constant of 56 in their bound. While it is possible that their analysis could be tightened up to reduce the constant somewhat, a
latency argument based approach is unlikely to yield optimal constant factors. 
 \cite<See the encyclopedia article on Min-Sum Set Cover by>[for a related discussion]{im:encyc16}.

Other bounds for Stochastic Submodular Cover have a dependence on quantities other than $Q$ and $\eta$.
In two related papers, Deshpande et al.~\citeyear{Deshpande:2016:AAS:2930058.2876506} and
Hellerstein and Kletenik~\citeyear{hellerstein:jair} 
showed that
Adaptive Greedy, with $\alpha=1$, achieves a bound of $k(\ln (Q/\eta)+1)$,
where $k$ is the number of states.
Their analysis used an LP formulation of the problem.
Recent independent work of Esfandiari et al.~\citeyear{Esfandiarietal19} on related problems shows that Adaptive Greedy with $\alpha=1$
achieves a bound close to $\ln (Qn/\eta)$.
In particular,
$c_{greedy} \leq (c_{opt} + 1)\ln (Qn/\eta)+ 1$
where $c_{greedy}$ is the expected cost of Adaptive Greedy,
and $c_{opt}$ is the expected cost of the optimal policy.
The analyses for these bounds, and for the 
$56(\ln (Q/\eta) + 1)$ bound of Im et al.,
can easily be extended to $\alpha \geq 1$, leading to an extra multiplicative factor of $\alpha$ in the bounds.

In this paper, we prove that the Adaptive Greedy algorithm yields an
$\alpha (\ln (Q/\eta) + 1)$ approximation bound for Stochastic Submodular Cover;
the bound is $\alpha H(Q)$, where $H(Q)$ is the $Q^{th}$ Harmonic number, when the utility function is integer valued. 
The tightness of the bound follows from the fact that it generalizes the $(\ln~m + 1)$ bound for classical Set Cover.  Unless $P = NP$, the classical Set Cover problem cannot be approximated to within a factor of  $(1-o(1))(\ln~ m)$~\cite{DinurSteurer,Feige98}. 

Our bound restores the bound originally claimed by Golovin and Krause 
for Stochastic Submodular Cover.  
It also generalizes the bound claimed by Liu et al.~\citeyear{Liu:2008:NAS:1376616.1376633} 
for their restricted version of Stochastic Set Cover.

\subsection{Related Results}

In this paper, we use the definition of the Stochastic Submodular Cover problem previously used by Deshpande et al.~\citeyear{Deshpande:2016:AAS:2930058.2876506} and by Hellerstein and Kletenik~\citeyear{hellerstein:jair}.  This definition is slightly more general than the original definition of Stochastic Submodular Cover given by Golovin and Krause~\citeyear{Golovin:2011:AST:2208436.2208448}. 
In Appendix~\ref{app:def}, we describe the differences in the definitions, and its relevance to the definition of the Weighted Stochastic Submodular Ranking problem studied by Im et al.~\citeyear{imetal}.  The analyses of Golovin and Krause, and of Im et al., still hold for the definition of Stochastic Submodular Cover used in this paper.  

In addition to defining Stochastic Submodular Cover, Golovin and Krause also defined a more general covering problem where item states are not necessarily independent
~\citeyear{Golovin:2011:AST:2208436.2208448,GolovinKrauseCorrected}. 
In this problem, the function $f$, together with the joint distribution on item states, are assumed to satisfy the properties of ``adaptive submodularity'' and ``adaptive monotonicity''.  These properties are defined in terms of the {\em expected} value of $f$ on random assignments of states to items.    
In their original paper, Golovin and Krause claimed the $\alpha(\ln (Q/\eta)+1)$ bound for this more general problem, and then applied it to Stochastic Submodular Cover. 
They later proved their corrected bound of
$\alpha(\ln(Q/\eta) + 1)^2$ for a version of their original problem
with somewhat stronger assumptions, and again 
applied the bound to Stochastic Submodular Cover.

In the previous section, we 
stated a bound for Stochastic Submodular Cover 
that is linear in $\ln(Qn/\eta)$,
implied by the work of
Esfandiari et al.~\citeyear{Esfandiarietal19}. 
In fact, Esfandiari et al.\ proved this bound
for the original version of the more general problem of Golovin and Krause,
assuming adaptive submodularity and adaptive monotonicity.

Our analysis of Stochastic Submodular Cover relies heavily on the assumption of independent item states, as do the analyses that led to the previous $56(\ln (Q/\eta) + 1)$ and $k(\ln (Q/\eta)+1)$ bounds. We leave open the
 question of whether an $\alpha(\ln (Q/\eta)+1)$ bound can be achieved for the more general problem of Golovin and Krause.

In their work on Stochastic Submodular Cover,
Deshpande et al.~\citeyear{Deshpande:2016:AAS:2930058.2876506} and Hellerstein and Kletenik~\citeyear{hellerstein:jair}
actually showed that
Adaptive Greedy with $\alpha=1$ achieves a bound of $k(\ln (R/\eta)+1)$.  Here $R$ is the largest increase in utility attainable from a single item, and thus is at most $Q$.    This is a generalization of the
$(\ln (R/\eta) + 1)$ bound of Wolsey for deterministic Submodular Cover.
Although the bound we prove in this paper is tight in terms of $Q/\eta$, we leave open the question of whether the $Q$ in our bound can be replaced by $R$.  For some problems, $R$ is much less than $Q$.

Hellerstein and Kletenik also proved an alternative bound of $(\ln (R_E/\eta_E)+ 1)$.  Here $R_E$ is the largest expected marginal increase in utility from a single item, and $\eta_E$ is the smallest expected non-zero marginal increase in utility from that same item.  This latter bound is incomparable to the  $(\ln (Q/\eta) + 1)$ bound shown in this paper, if items can have non-zero utility in one state, and zero utility in another.  

To achieve the bound in this paper, we use an amortized approach.  
Liu et al.~\citeyear{Liu:2008:NAS:1376616.1376633} also used an amortized approach in their incorrect proof, but the approach here is different.  A major difference is that Liu et al.\ only considered the Stochastic Set Cover problem, and their approach relied on charging costs to elements of the ground set as they were covered.  In our problem, Stochastic Submodular Cover, costs cannot be charged in this way, since there is no ground set, but only an abstract submodular utility function.  This makes amortization more challenging.  

An analysis due to Wan et al.~\citeyear{WanDPW10} for a deterministic generalization of Submodular Cover used an amortized approach with an abstract submodular utility function, but it did not need to address the challenges arising from the stochastic nature of our problem and the branching in the adaptive ordering. 

We note that Deshpande et al.\
also introduced a dual greedy algorithm for Stochastic Submodular Cover, and proved that it achieves a bound that is qualitatively different and incomparable to the bounds given above.  The bound generalizes the dual greedy approximation bound of Hochbaum~\citeyear{hochbaumDualGreedy}  for the classical Set Cover problem, and Fujito's extension of it to Submodular Cover  \cite<cf.>{fujitoSurvey}.  These bounds all assume that the dual greedy algorithm can find the  exactly optimal dual greedy choice at every step.  It is not clear how to extend these results to approximately optimal dual greedy choices.  
\label{sec:prelim}
\section{Preliminaries}

{\it For the aid of the reader, a table of frequently used notation is presented in Appendix~\ref{sec:symbols}.} 

We now present a formal definition of the Stochastic Submodular Cover problem and introduce associated notation.  

Let $E=\{e_1, \ldots, e_n\}$ be a finite set of {\em items}.
Let $O = \{o_1, \ldots, o_k\}$ be a finite set of {\em states}.
A {\em realization} is a function $\varphi:E \rightarrow O$, assigning state $\varphi(e)$ to item $e$.
A {\em subrealization} is a function $\psi:E \rightarrow O \cup \{*\}$ which is a partial assignment of states to items (we assume $* \not\in O$).
If  $\psi(e) \in O$ then $\psi(e)$ is the state assigned to $e$, while if $\psi(e) = *$, then the state of $e$ is unassigned.
We define 
\begin{gather*}
dom(\psi) \triangleq \{e \in E \vert \psi(e) \in O\}.  
\end{gather*}
Realization $\varphi$ has a corresponding relation 
\begin{gather*}
rel(\varphi) \triangleq \{(e,o) \in E \times O \vert \varphi(e) = o\}.
\end{gather*}
Subrealization $\psi$ has a corresponding relation $rel(\psi) \triangleq \{(e,o) \in E \times O \vert e \in dom(\psi) \wedge \psi(e) = o\}$.

Given subrealizations $\psi$ and $\psi'$, we write $\psi \subset \psi'$ to denote that $rel(\psi) \subset rel(\psi')$ and $\psi \subseteq \psi'$ to denote that $rel(\psi) \subseteq rel(\psi')$.  When $\psi \subseteq \psi'$ we call $\psi'$ an {\em extension} of $\psi$.

Given a subrealization $\psi$ and a pair $(e,o) \in E \times O$ where $e \not\in dom(\psi)$, we use $\psi \cup \{(e,o)\}$ to denote the extension $\psi'$ of $\psi$ such that $rel(\psi') = rel(\psi) \cup \{(e,o)\}$.  

Let $\mathbb{D}$ be a distribution 
whose domain is the set of
all realizations
 $\varphi:E \rightarrow O$.  Let $\Phi$ be a random variable distributed according to distribution $\mathbb{D}$.
Thus $\Phi$ is a random assignment of states to items.

In what follows, all probabilities and expectations are with respect to distribution $\mathbb{D}$ for $\Phi$.

\begin{assumption}[Independence]
  \label{assum:independence}
    We assume that in distribution $\mathbb{D}$, item states are independent, i.e., for all realizations $\varphi:E \rightarrow O$,
  \begin{gather*}
 P[\Phi = \varphi] = \prod_{e \in E} P[\Phi(e)=\varphi(e)].
  \end{gather*}
\end{assumption}

Let $\mathcal{F} = \{f^{\varphi} \vert \varphi:E \rightarrow O \}$ be a
family of utility functions $f^{\varphi}:2^E \rightarrow \mathbb{R}_{\geq 0}$, whose elements $f^{\varphi}$ have a one-to-one correspondence with the set of realizations $\varphi$.  
We assume that the functions $f^{\varphi}$ in $\mathcal{F}$ have the following two properties:

\begin{assumption}[Pointwise Polymatroid]
  \label{assum:pointwise}
  Every function in $\mathcal{F}$ is a polymatroid function.
\end{assumption}

\begin{assumption}[Sufficiency]
  \label{assum:sufficiency}
  Given $E' \subseteq E$ and
  realizations $\varphi_1$, $\varphi_2$ such that 
  $\varphi_1(e) = \varphi_2(e)$ for all $e \in E'$, we
  have
  $f^{\varphi_1}(E') = f^{\varphi_2}(E')$.
\end{assumption}

Assumption~\ref{assum:sufficiency} (Sufficiency)
says that the value of $f^{\varphi}$ on any subset of items $E' \subset E$
depends only on the states that $\varphi$ assigns to
items in $E'$, and not to the states it assigns to
items outside of $E'$.

Define $f$ to be the mapping from  subrealizations $\psi$ to $\mathbb{R}_{\geq 0}$ such that
$f(\psi) = f^{\varphi}(dom(\psi))$, for all realizations
$\varphi$ such that $\psi \subseteq \varphi$.
By the Sufficiency Assumption, $f$ is well-defined.

\begin{assumption}[Coverability]
\label{assum:coverability}
There is a value $Q \in \mathbb{R}_{>0}$ such that for all realizations $\varphi$, $f(\varphi) = Q$.
\end{assumption}

We call $Q$ the {\em goal value} of $f$.  We say that subrealization $\psi$ is a {\em cover} of $f$ if $f(\psi) = Q$.

Relative to $\psi$,
the marginal increase in utility due to observing $e \in E \setminus dom(\psi)$ in state $o$, is 
\begin{gather*}
f_{\psi}(e,o) \triangleq f(\psi \cup \{(e,o)\})  - f(\psi)
\end{gather*}
We use $F_{\psi}(e)$ to denote the random marginal increase in utility from $e \in E \setminus dom(\psi)$,
relative to $\psi$, that will result from observing the state $\Phi(e)$ of $e$.  That is,
\begin{gather*}
F_{\psi}(e) \triangleq f_{\psi}(e,\Phi(e))
\end{gather*}

An {\em adaptive covering policy} $\pi$ 
for $f$ is a function which maps every subrealization $\psi$ such that $f(\psi) < Q$
to an item
$\pi(\psi) \in E \setminus dom(\psi)$.
\footnote{
We consider only deterministic policies, but our approximation bound for Adaptive Greedy also holds relative to a randomized policy $r$ with minimum expected cost.  
 Under the standard assumptions on the sources of randomness, executing $r$ is equivalent to first choosing  
a policy $\pi$ at random from a distribution $D$ over deterministic policies, such that $\pi$ is independent of $\Phi$, and then executing $\pi$. Since $r$ is optimal, every policy in the support of $D$ has the same expected cost as $r$.
Since our approximation bound holds with respect to each $\pi$ in the support of $D$, it also holds with respect to $r$.}


Policy $\pi$ specifies a sequential order in which to select the items of a cover for $f$.  
The order is adaptive, meaning that the choice of the next item depends on the states of the already selected items.  

 The selection of items proceeds in {\em iterations}. 
In each iteration, an item $e$ is selected and its state is observed.  Each item can only be selected once.  
As the iterations proceed, the set of selected items, and their states, can be represented by a subrealization $\psi$: $dom(\psi)$ is the set of selected items, and for each $e$ in that set, $\psi(e)$ is its observed state.  
Initially $\psi$ is such that $rel(\psi)=\emptyset$.
Policy $\pi$ specifies the item to select in each iteration, namely item $e=\pi(\psi)$.  Selection of items, and observations of their states (with associated updating of $\psi)$, continues until $\psi$ is a cover for $f$. 
Executing policy $\pi$ on realization $\varphi$ means following this procedure when the observed state of each selected item $e$ is $\varphi(e)$.

\begin{figure}[ht]
    \centering
    \includegraphics[width=0.75\textwidth]{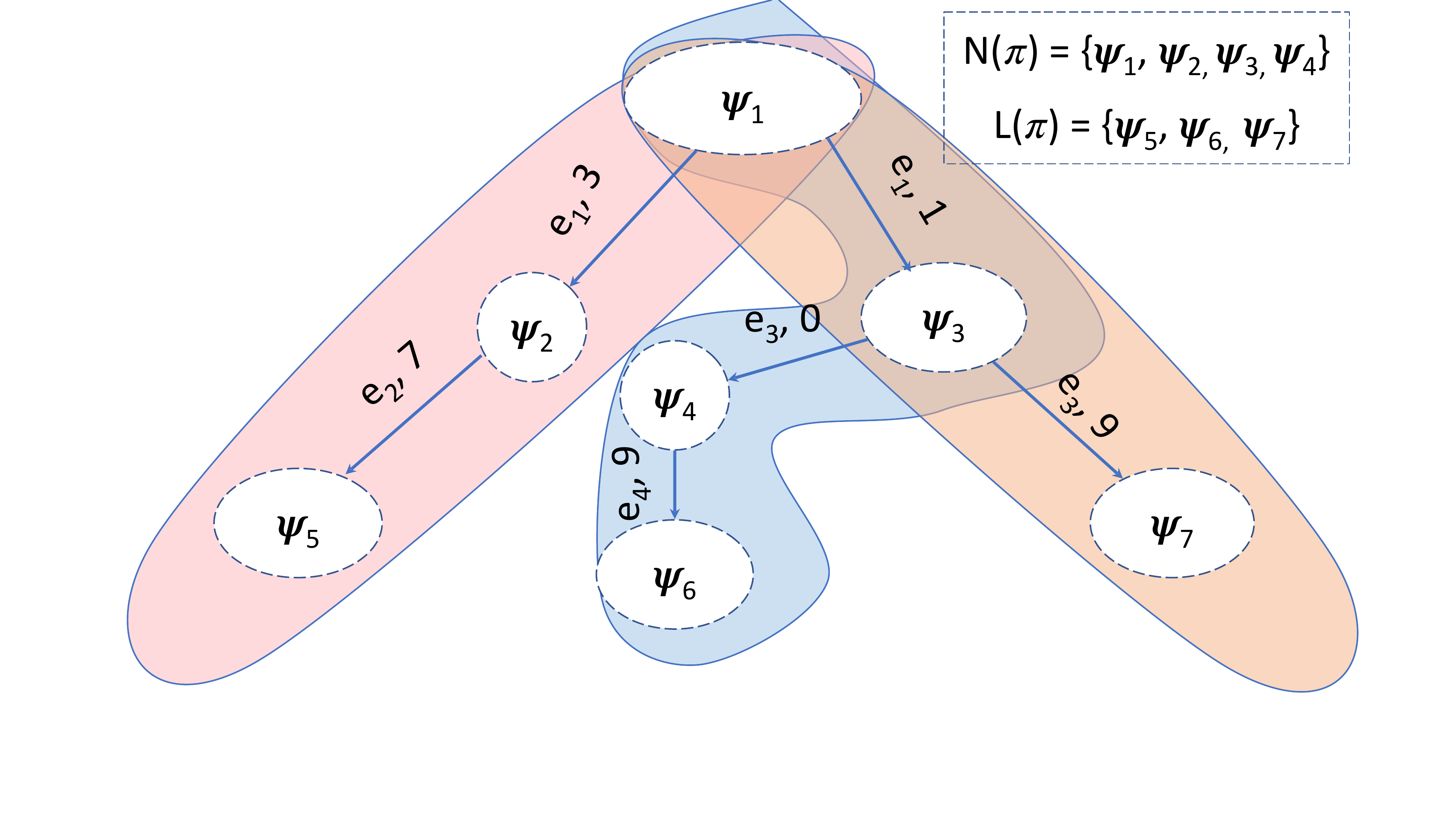}
      \caption{Tree representation of a policy $\pi$ with $Q = 10$.  
       Annotation $(e,\delta)$ on edge $(\psi_i,\psi_j)$ indicates that $e$ is selected at node $\psi_i$
       and $\delta = f(\psi_j) - f(\psi_i)$.
      }
      \label{fig:tree}
  \end{figure}

The adaptive sequential order specified by policy $\pi$ can be represented by a decision tree $T(\pi)$, whose nodes are identified with subrealizations $\psi$. 
The root of the tree corresponds to the initial iteration, identified with realization $\psi$ where $rel(\psi) = \emptyset$. The leaves are identified with the possible covers of $f$ constructed by the policy.  
An example tree is shown in Figure~\ref{fig:tree}.

Each non-leaf node $\psi$ of the tree corresponds to an iteration where $\psi$ represents the states of the items selected so far.
The children of node $\psi$ correspond to the possible updated values for $\psi$ at the end of the iteration, after the state of selected item $\pi(\psi)$ is observed.  Letting $e = \pi(\psi)$, 
for each $o \in O$ there is a child node identified with $\psi' = \psi \cup \{(e,o)\}$, if there is a non-zero probability of observing state $o$ for
item $e$ under distribution $\mathbb{D}$ (i.e., if
$P[\Phi(e)= o] > 0$). 
In Figure~\ref{fig:tree}, the edge from $\psi$ to child $\psi'$ is labeled with the selected item $e$, and with the associated increase in utility $f(\psi') - f(\psi)$.

By the Independence Assumption, for subrealization $\psi$ we have
\begin{gather*}
    \forall e \in E \setminus dom(\psi): ~~~\E[F_{\psi}(e) \vert \psi \subset \Phi] = \E[F_{\psi}(e)]
\end{gather*}

Let $L(\pi)$ be the set of leaf nodes of $T(\pi)$ and let $N(\pi)$ be the set of non-leaf nodes.
If subrealization $\psi$ is a non-leaf node then
$f(\psi) < Q$, and if it is a leaf node then $f(\psi) = Q$.

The execution of $\pi$ on random realization $\Phi$ corresponds to a random walk on $T(\pi)$ which starts at the root and
proceeds downwards towards a random leaf of the tree.
We regard the walk as visiting not just a sequence of nodes, but also the associated sequence of subrealizations. 
Formally, the associated sequence of subrealizations is
$\psi_1, \psi_2, \ldots, \psi_k$ for some $k > 0$ where 
$f(\psi_1) = \emptyset$, $f(\psi_k) = Q$, and for $i=1, \ldots,  k-1$, $f(\psi_i) < Q$ and $\psi_{i+1} = \psi_i \cup \{(\pi(\psi_i),\Phi(\pi(\psi_i))\}$.  
The final subrealization in this sequence, $\psi_k$, is identified with the leaf at the end of the path in $T(\pi)$ corresponding to the sequence.  
We refer to the  final subrealization of the sequence as the {\em cover constructed by executing $\pi$ on $\Phi$} and define the random variable
\begin{gather*}
    \Psi(\pi) \triangleq \text{the cover constructed by executing } \pi \text{ on random realization } \Phi.
    \nonumber
\end{gather*}

For $\psi \in N(\pi)$ and
$e = \pi(\psi)$, 
$\E[F_{\psi}(e)]$
is the expected increase in utility at node $\psi$ of $T(\pi)$, immediately prior to observing the state of $e$.

Consider the execution of $\pi$ on random realization $\Phi$.  If $\psi \in N(\pi)$ is visited during this execution, then clearly $\psi \subset \Psi(\pi)$ and $\psi \subset \Phi$.  If $\psi$ is not visited, then there must be an ancestor $\psi'$ of $\psi$ 
such that the item $\pi(\psi)$ chosen at node $\psi$ is assigned to different states by $\psi$ and by $\Psi(\pi)$.
Thus the following  conditions are equivalent for $\psi \in N(\pi)$: (1) $\psi$ is visited when executing $\pi$ on $\Phi$,  (2) $\psi \subset \Phi$, and (3) $\psi \subset \Psi(\pi)$.  

We define $S(\pi)$ to be the random
set of items selected by $\pi$ when executed on random realization $\Phi$.  That is,
\begin{gather*}
    S(\pi) \triangleq dom(\Psi(\pi))  
\end{gather*}

We define
\begin{gather*}
  C(\pi) \triangleq \sum_{e \in S(\pi)} c(e)   
\end{gather*}
Thus $C(\pi)$ is a random variable whose value is the sum of the costs incurred by executing $\pi$ on  $\Phi$.

Formally, the inputs to the Stochastic Submodular Cover problem, consistent with the definitions and assumptions above, are as follows: 
(1) the set $E=\{e_1, \ldots, e_n\}$ of items, 
(2) the value of $P[\Phi(e) = o]$ under distribution $\mathbb{D}$, for each $e \in E$ and $o \in O$,
(3) a cost function $c:E \rightarrow \mathbb{R}_{>0}$, and 
(4) an oracle that takes as input any subrealization $\psi$, and outputs the value $f(\psi)$.
The problem is to find an  adaptive covering policy $\pi$ for $f$
with minimum expected cost, where the expectation is with respect to the distribution $\mathbb{D}$ of $\Phi$:
\begin{gather*}
  \min_{\pi}~~\E[C(\pi)]
\end{gather*}
The approximation bound we prove in this paper depends on the quantity $\eta$,
which is the minimum gap between $Q$ and $f(\psi)$ for any subrealization $\psi$ where $f(\psi) < Q$.  Formally:
   \begin{gather*}
   \eta \triangleq \min\{Q-f(\psi) \vert \psi:E \rightarrow O \cup \{*\} \mbox{ and } f(\psi) < Q \}
   \end{gather*}
 When $f$ is integer valued, $\eta \geq 1$.

In our analysis, we will sometimes use the following intuitive perspective.    
Think of the
execution of an adaptive covering policy $\pi$ on $\Phi$ in terms of driving a car along a road of length $Q$, with stops along the way. 
The start of the execution corresponds to the start of the drive.  
During the execution, when an item $e$ is selected and its state is observed, there is a marginal increase $\delta$ in utility, and cost $c(e)$ is incurred.    
This corresponds to driving a further distance of $\delta$ along the road, at a cost of $c(e)$, before stopping again.  At the end of execution, the sum of the increases in utility is $Q$, which corresponds to having traveled the entire length $Q$ of the road.

\section{Adaptive Greedy Policy}
\label{sec:sgreedy}

The Adaptive Greedy algorithm of Golovin and Krause~\citeyear{Golovin:2011:AST:2208436.2208448,GolovinKrauseCorrected}, when used to solve Stochastic Submodular Cover, executes the {\em adaptive greedy policy} defined as follows.



\bigskip

  Define the \textit{unit price} of an
  item $e \in E \setminus dom(\psi)$ with respect to subrealization $\psi$, denoted by $\upsilon(e \vert \psi)$, as follows:
  
  \bigskip
  \begin{gather}
     \upsilon(e \vert \psi) \triangleq
    \frac{c(e)}{\E[F_{\psi}(e) \vert \psi \subset \Phi]} = \frac{c(e)}{\E[F_{\psi}(e)]} \label{eqn:unitprice}
  \end{gather}

\bigskip
The unit price is considered to be infinite if $\E[f_{\psi}(e,\Phi(e))] = 0$.

We denote the adaptive greedy policy by $\sigma$.  
The adaptive greedy policy with $\alpha = 1$ is the adaptive covering policy that, for
subrealization $\psi$ such that $f(\psi) < Q$, 
selects the item $e$ with smallest unit price, so 
%
$\sigma(\psi) = \arg\min_{e \in E  \setminus dom(\psi)} \upsilon(e \vert \psi)$
{\noindent with} ties broken deterministically.

In some applications, though, finding the item with smallest unit price can be a computationally hard problem.
We may instead solve the problem
only approximately, with an approximation ratio $\alpha$. 

More generally, then, for fixed $\alpha \geq 1$, the adaptive greedy policy $\sigma$
selects some element $\sigma(\psi)$ such that
\begin{gather}
  \upsilon(\sigma(\psi) \vert \psi)
  \leq \alpha \left(\min_{e \in E \setminus dom(\psi)} \upsilon(e \vert \psi) \right) 
\end{gather}

\bigskip
We use $e_{\psi}$ to denote the item $\sigma(\psi)$.
We use
$\upsilon_{\psi}$ to denote the quantity $\upsilon(e_{\psi} \vert \psi)$.  Thus

\begin{gather}
  \upsilon_{\psi} \triangleq \upsilon(e_{\psi} \vert \psi) \leq \alpha \left ( \min_{e \in E\setminus dom(\psi)} \upsilon(e \vert \psi) \right )
  \label{eqn:greedyrate}
\end{gather}

\section{Comparison of $\pi^*$ and $\sigma$}

Let $\pi^*$ be an optimal adaptive covering policy for an instance of Stochastic Submodular Cover.  We will prove an approximation bound that compares $\E[C(\sigma)]$, the expected cost of the adaptive greedy policy, to $\E[C(\pi^*)]$, the expected cost of the optimal policy.   To accomplish this,
we introduce different ways of accounting for the costs incurred while executing policies $\pi^*$
and $\sigma$ on the same random realization $\Phi$.

Given item $e$ and policy $\pi$, let $\pi_{-1}[e]$ denote the \textit{inverse image}
of $\pi$ under $\{e\}$: this is the set of all subrealizations corresponding to nodes in $N(\pi)$ at which item $e$ is selected.

Consider the execution of $\pi^*$ and $\sigma$ on the same random realization $\Phi$.  
The two executions give rise to values for random variables $S(\pi^*)$ (the items selected by policy $\pi^*$), $S(\sigma)$ (the items selected by policy $\sigma$), and $\Psi(\sigma)$ (the cover constructed by $\sigma$).
We define the {\em revenue} collected by $\pi^*$ and $\sigma$,
in their executions on realization $\Phi$,
in terms of these random variables.   

\subsection*{Revenue collection by $\pi^*$}
\label{perspective:pi*revenue}

Policy $\pi^*$ collects its revenue from the items $e \in E$.
For all $e \in E$,
we define the revenue $\mu(e)$ that $\pi^*$ collects from $e$ as follows:
 
 \boxedeq{eqn:mu}{
     \mu(e) \triangleq
      \begin{cases}
        0 & \text{if } e \not\in S(\pi^*) \\
        c(e) & \text{if } e \in S(\pi^*) \setminus S(\sigma) \\
        \upsilon_{\psi}  F_{\psi}(e) & \text{if } e \in S(\pi^*) \cap S(\sigma), \text{ where } \psi \text{ is the}\\ & ~~~~~~\text{subrealization in } \sigma_{-1}[e] \text{ s.t. } \psi \subset \Psi(\sigma)
      \end{cases} }
  
  \bigskip
  {\noindent That is, if} $e$ was not selected by $\pi^*$, it does not contribute any revenue. If $e$ was selected
  by $\pi^*$ but not by $\sigma$, then its revenue equals $c(e)$. Otherwise,
  let $\psi$ be the  subrealization visited by $\sigma$ where $e$ was selected: the revenue in this case is defined to be the
  product of the unit price $\upsilon_{\psi}$ and the marginal increase in utility obtained by $\sigma$ when it
  observed $e$ in state $\Phi(e)$.
  
  The total revenue collected by $\pi^*$ is
  \begin{gather}
    \mu^* \triangleq \sum_{e \in E} \mu(e) \label{eqn:mu*}
  \end{gather}

\subsection*{Revenue collection by $\sigma$}

  \label{perspective:sigmarevenue}
  Policy $\sigma$ also 
 collects revenue when executed on $\Phi$. 
  The revenue in this case is defined as follows.  
  Let $\rho_1, \rho_2, \ldots, \rho_{\cardinality{N(\sigma)}}$ be all the subrealizations in
  $N(\sigma)$ indexed in non-decreasing order of their utilities so
  that $f(\rho_1) \leq f(\rho_2) \leq \ldots \leq f(\rho_{|N(\sigma)|})$
  (ties are broken arbitrarily).
  We treat each $\rho_i$ as a \textit{marker} located at
  point $f(\rho_i)$ on the interval $[0, Q]$ (corresponding to the road of length $Q$ in our driving formulation).
  Let $\rho_{\cardinality{N(\sigma)} + 1}$ be an extra marker placed at point $Q$.
  Define $f(\rho_{\cardinality{N(\sigma)} + 1}) \triangleq Q$, and
  for $i \in \{ 1, \ldots, \cardinality{N(\sigma)}\}$ define $\epsilon_i \triangleq f(\rho_{i+1}) - f(\rho_i)$.
  
  Observe that $\epsilon_i$ is the length of the $i^{th}$ subinterval, which lies between $\rho_i$ and $\rho_{i+1}$.  Also, if $\psi$ is the subrealization corresponding to a node in $T(\sigma)$, then $f(\psi)$ is on the boundary of one of these subintervals.  We have 
  \begin{gather}
    \sum_{i=1}^{\cardinality{N(\sigma)}} \epsilon_i = Q \label{eqn:sumepsilons} 
    \end{gather}
    \begin{gather}
     \epsilon_{\cardinality{N(\sigma)}} 
     \geq \eta 
     \label{eqn:lastepsilon}
  \end{gather}
  
\smallskip  
{\noindent where} \eqref{eqn:lastepsilon} follows because $f(\rho_{\cardinality{N(\sigma)}}) < Q$ and $f(\rho_{\cardinality{N(\sigma)}+1}) = Q$.
\bigskip
  
  Policy $\sigma$ collects its revenue from the markers $\rho_i$.   Viewing the execution of $\sigma$ in terms of the drive along the road of length $Q$, revenue is collected from marker $\rho_i$ when the car reaches a distance of $f(\rho_i)$ from the start.

 We now describe the details of how $\sigma$ collects revenue from $\rho_i$.
  Let $R_i$ be the set of markers in the interval $[0, f(\rho_i)]$. 
  Define:
  \smallskip
  \begin{itemize}
  \item[] $\psi \leadsto i \triangleq$ 
  the event that, in the execution of $\sigma$ on $\Phi$, $\psi$ was the last visited subrealization in $R_i$ 
  \end{itemize}
  \smallskip
  That is, $\psi \leadsto i$ means that among all subrealizations that were visited when executing $\sigma$ on $\Phi$, $\psi$ is the last one visited whose utility did not exceed $f(\rho_i)$.
  
  Note that if $f(\rho_i) = f(\rho_{i+1})$,  then $R_i$ and $R_{i+1}$ are equal and contain both $\rho_i$ and $\rho_{i+1}$.
  For each realization $\Phi$ and $i \in \{1, \ldots, |N(\sigma)|\}$
  there is a unique $\psi \in R_i$ satisfying $\psi \leadsto i$.

  Before completing the description of the revenue collection process for $\sigma$, we give an example to illustrate the above definitions.
  
\begin{figure}[ht]
    \centering
      \includegraphics[width=\textwidth]{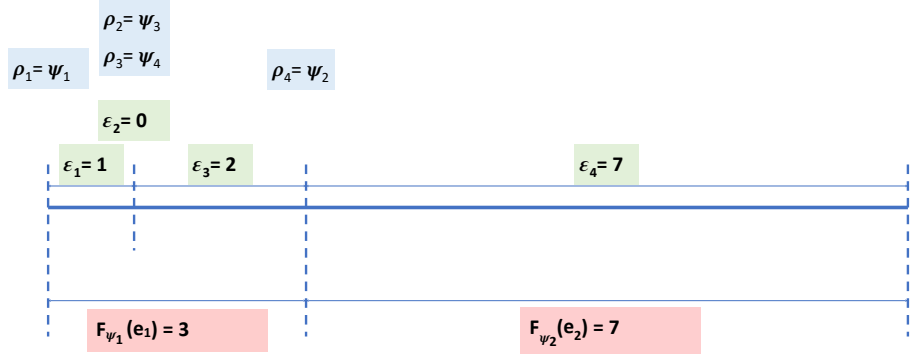}
      \caption{Revenue collection when $\sigma$ is the policy $\pi$ in Figure~\ref{fig:tree}.}  
     
      \label{fig:markers}
  \end{figure}

  \bigskip
  {\bf \noindent Example:} {\it Suppose that $\sigma$ is the policy $\pi$ whose tree is shown in Figure~\ref{fig:tree}.
  Here $Q=10$, $f(\psi_1) = 0$, $f(\psi_2) = 3$, and $f(\psi_3) = f(\psi_4) = 1$.
  Breaking ties by increasing index, we have markers $\rho_1 = \psi_1$, $\rho_2 = \psi_3$, $\rho_3 = \psi_4$, and $\rho_4 = \psi_2$, and we place them at the points $f(\rho_1)=0$, $f(\rho_2)=1$, $f(\rho_3)=1$, and $f(\rho_4)=3$ along the interval $[0,10]$. 
  This is shown graphically in Figure~\ref{fig:markers}, along with the associated values of the $\epsilon_i$.  The additional marker $\rho_5$,
   which is not shown in the figure, is located at the right end of the interval, at  $f(\rho_5)=10$.
  Suppose that the execution of $\sigma$ on $\Phi$ terminates with subrealization $\psi_5$, so  $\Psi(\sigma) = \psi_5$.  In this case 
  the subrealizations in $N(\sigma)$ visited during the execution are $\psi_1$ and $\psi_2$; there is an increase of 3 units of utility after observing the state of $e_1$, followed by an increase of 7 units after observing the state of $e_2$.
  Further, $\psi_1 \leadsto i$ for $i=1, \ldots, 3$, while $\psi_2 \leadsto i$ for $i=4$.  The marginal increases in utility for this case are shown at the bottom of Figure~\ref{fig:markers}.  
  
  If, however, $\Psi(\sigma)=\psi_6$, then after $\psi_1$, policy $\sigma$ visits $\psi_3$ and $\psi_4$ before reaching a leaf.  In this case, $\psi_1 \leadsto i$ for $i=1$, and $\psi_4 \leadsto i$ for $i=2,3,4$.}
  
  \bigskip
  
  Policy $\sigma$ collects a random amount of revenue $\lambda_i$ from marker $\rho_i$ as follows:
  
  \boxedeq{eqn:lambdai}{
   \lambda_i \triangleq \upsilon_{\psi}\epsilon_i, \text{ where  }\psi \in R_i \text{ is such that } \psi \leadsto i }
  
  \bigskip
  By the monotonicity property of the Pointwise Polymatroid Assumption, the event $\psi \leadsto i$ implies that
  there is \text{no} subrealization $\psi' \in N(\sigma)$ such that
  $\psi' \subset \Psi(\sigma)$ and $f(\psi') \in (f(\psi), f(\rho_i)]$.  This implies the following fact.
  
  \bigskip
  \begin{fact}
  \label{fact:lambdapsi}
  \begin{gather}
    \forall \psi \in N(\sigma) \text{ such that } \psi \subset \Psi(\sigma):~\sum_{i \text{ s.t. } \psi \leadsto i} \lambda_i
    ~~{=}~~ \upsilon_{\psi} \sum_{i \text{ s.t. } \psi \leadsto i} \epsilon_i
    ~~=~~ \upsilon_{\psi} F_{\psi}(e_{\psi})
    \nonumber
  \end{gather}
  \end{fact}
  
  \subsection*{Hybrid policy}
  We also define a hybrid policy that 
  combines $\sigma$
  and $\pi^*$ in a careful manner. In particular, let $\psi$ be a subrealization such that $\psi \in N(\sigma)$. 
  We define the hybrid policy $\pi^{\psi}$ whose execution on a random realization $\Phi$ consists of two stages.
  In the first stage, the execution of $\pi^{\psi}$ on $\Phi$ follows the same sequence as $\sigma$ until
  it either visits $\psi$ or terminates.
  As soon as it visits $\psi$, $\pi^{\psi}$  enters its second stage, during which it executes policy $\pi^*$, starting from the root of $T(\pi^*)$ (i.e., as if it has not yet selected any items).  However,
  during the execution of $\pi^*$, $\pi^{\psi}$ does not re-select an item $e$ if it was already selected during the first stage.  Instead, $\pi^{\psi}$ uses the previously observed state of $e$ (which is $\psi(e)$) and proceeds with the execution of $\pi^*$ as if $\pi^*$ had observed that state for $e$.\footnote{Technically, our description of $\pi^{\psi}$ should also specify the value of
  $\pi^{\psi}(\psi')$ for subrealizations $\psi'$ that do not appear as nodes in its decision tree.  
  These values do not affect execution of $\pi^{\psi}$ and can be assigned arbitrarily.}
  
  Note that when $\pi^{\psi}$ is executed on a random realization $\Phi$ such that 
  $\psi \not\subset \Psi(\sigma)$ (i.e., $\sigma$ does not visit $\psi$ when executed on $\Phi$),
  then $\pi^{\psi}$ selects the same items that would have been selected by executing $\sigma$ on $\Phi$.  When $\pi^{\psi}$ is executed on a realization $\Phi$ such that $\psi \subset \Psi(\sigma)$, then the items it selects include all items that would have been selected by executing $\pi^*$ on $\Phi$.  It follows that $\pi^{\psi}$ is an adaptive covering policy.
  
  Because $\sigma$ selects item $e_{\psi}$ at node $\psi$ of $T(\sigma)$, it follows that $e_{\psi} \not\in dom(\psi)$.

Further, by the definition of the hybrid policy, we have the following fact.

\begin{fact}
\label{fact:opttohybrid}
\begin{gather}
\forall \psi \in N(\sigma) \text{ such that }\psi \subset \Psi(\sigma),~\forall e \in E \setminus dom(\psi): ~e \in S(\pi^*) \text{ iff } e \in S(\pi^{\psi})
\end{gather}
\end{fact}  
  
  We define $\delta^+_{\psi}(e)$ to be the marginal utility obtained by
  $\pi^{\psi}$ when $e$ is selected, if $e$ is selected in the second stage; 
  if $e$ is not selected by $\pi^{\psi}$ in the second stage, this value is $0$.
%
  
  Formally:
  \boxedeq{eqn:deltaplus}{
     \delta^+_{\psi}(e) \triangleq
      \begin{cases}
        F_{\psi'}(e) & \text{ if } \exists \psi' \in
         \pi^{\psi}_{-1}[e] \text{ s.t. } \psi \subseteq \psi',~ \psi' \subset \Psi(\pi^{\psi}) \\
        0 & \text{ otherwise} 
      \end{cases} 
      }

   \section{Approximation Analysis: Overview}
   \label{sec:overview}
   
Recall that $\eta$ is equal to the minimum value of $Q-f(\psi)$ over all subrealizations $\psi$ such that $f(\psi) < Q$.
The goal of our approximation analysis is to prove the following theorem:
 
 \begin{theorem}
\label{theo:greedy}
The adaptive greedy policy $\sigma$
achieves an approximation bound of
$\alpha\kappa$ for Stochastic Submodular Cover, where $\kappa = \ln (Q/\eta)+1$
when the utility function $f$ is real valued and $\kappa = H(Q)$ when the utility function is integer valued.
That is, 
$\E[C(\sigma)] \leq \alpha \kappa \E[C(\pi^*)]$.
\end{theorem}
  
 We present the proof of the theorem in the following section.  It is based on a sequence of lemmas.
 We first prove that the revenue collecting procedures defined above correctly account for the expected cost incurred by their associated algorithms. 
 Specifically, in Lemma~\ref{lemma:dsigma=lambda}, we prove 
 $\E[C(\sigma)] = \sum_{i=1}^{\cardinality{N(\sigma)}} \E[\lambda_i]$, meaning that the expected cost of the greedy policy $\sigma$ is equal to the expected revenue it collects from the markers $\lambda_i$.
 In Lemma~\ref{lemma:d*>=mu*} we prove $\E[C(\pi^*)] = \E[\mu^*]$, meaning that
 the expected cost of the optimal policy $\pi^*$ is equal to the revenue it collects from the items $e$.
  
  We then focus on comparing  $\E[\mu^*]$
  to 
  $\sum_{i=1}^{\cardinality{N(\sigma)}} \E[\lambda_i]$.
  To make this comparison,
  consider $\lambda_i$, the revenue collected by $\sigma$ at marker $\rho_i$. By the definition of $\lambda_i$, its value depends on the sequence of subrealizations $
  \psi$ visited by $\sigma$ when executed on $\Phi$.  In particular, it depends on
  which $\psi$ in that sequence satisfies 
  $\psi \leadsto i$
  (i.e., which
  is the last  subrealization in the sequence that satisfies $f(\psi) \leq f(\rho_i))$.
  For each $i$,
  we partition the sample space according to which $\psi$ satisfies $\psi \leadsto i$.  In Lemmas
  \ref{lemma:deltapluslem} and~\ref{lemma:conditionallambdai}, we prove bounds relating the quantities $\E[\lambda_i \vert \psi \leadsto i]$, $\E[\mu(e) \vert \psi \leadsto i]$, and $\E[\delta^+_{\psi}(e) \vert \psi \leadsto i]$.
  
  Lemma~\ref{lemma:deltapluslem}
  shows that for all items $e \not\in dom(\psi)$,
  $\alpha\E[\mu(e) \vert \psi \leadsto i]
    \geq
    \E[\delta^+_{\psi}(e) \vert \psi \leadsto i] \upsilon_{\psi}$. 
 Recall that $\delta^+_{\psi}(e)$ is the expected increase in utility from $e$ in the second stage of $\pi^{\psi}$ (after the switch to $\pi^*$), and $\upsilon_{\psi}$ is the unit price paid by $\sigma$ when it makes its greedy choice at node $\psi$ of tree $T(\sigma)$.
  
  Lemma~\ref{lemma:deltapluslem} is used to prove
  Lemma~\ref{lemma:conditionallambdai}, which
  states that 
  $\forall i \in \{1, \ldots, \cardinality{N(\sigma)} \},~ \E[\lambda_i \vert \psi \leadsto i]
    \leq \frac{\alpha\E[\mu^* \vert \psi \leadsto i]}{Q - f(\rho_i)}\epsilon_i$.
  For some intuition behind this statement, consider the execution of the greedy policy, and the corresponding drive along the road of length $Q$, conditional on $\psi \leadsto i$.  After the car reaches marker $\rho_i$ (has traveled a distance of $f(\rho_i)$), it travels an additional distance $\epsilon_i$ until it reaches the next marker $\rho_{i+1}$.  If we view $\lambda_i$ as the revenue paid to finance this travel, then the expected revenue collected by $\sigma$, per unit distance, in traveling from $\rho_i$ to $\rho_{i+1}$, is $\E[\lambda_i \vert \psi \leadsto i]/\epsilon_i$.
  Consider this ratio to be the rate paid by $\sigma$ in traveling from $\rho_i$ to $\rho_{i+1}$.  
  Then consider what happens in the drive corresponding to policy $\pi^*$ when it reaches marker $\rho_i$. 
  At marker $\rho_i$, the remaining distance to be traveled is $Q - f(\rho_i)$.   Since the total revenue collected by $\pi^*$ is $\mu^*$, it will collect at most that much revenue in traveling the remaining distance $Q - f(\rho_i)$. 
Thus $\E[\mu^* |\psi \leadsto i]$ is an upper bound on the expected revenue collected by $\pi^*$ while traveling the remaining distance,
and the ratio
 $\E[\mu^* \vert \psi \leadsto i]/(Q - f(\rho))$ is an upper bound on the rate paid by $\pi^*$ during that travel.  Lemma~\ref{lemma:conditionallambdai} says that this ratio (times $\alpha$) is an upper bound on the rate paid by $\sigma$ in traveling from marker $\rho_i$ to $\rho_{i+1}$. 
  
  The next lemma, Lemma~\ref{lemma:unconditionallambdai}, 
  states that
  $\forall i \in \{1, \ldots, \cardinality{N(\sigma)} \},~ \E[\lambda_i] \leq \alpha \E[C(\pi^*)] \frac{\epsilon_i}{Q - f(\rho_i)}$.
  It is the same as Lemma~\ref{lemma:conditionallambdai}, except that the conditioning on $\psi \leadsto i$ is removed, and we use the fact that $\E[\mu^*] = \E[C(\pi^*)]$.
  Having removed the conditioning, we no longer need to worry about the branching in the decision tree for $\sigma$, and can simply think of $\sigma$ as deterministically traveling step by step from each marker $\rho_i$ to the next one, at a cost of $\E[\lambda_i]$ for the step. 
  
  The rest of the analysis is done in a standard way and
  leads to the desired approximation bound for the adaptive greedy policy:
  $\E[C(\sigma)] \leq \alpha \E[C(\pi^*)] (\ln(Q/\eta) + 1)$ (and 
  $\E[C(\sigma)] \leq \alpha \E[C(\pi^*)] H(Q)$ when $f$ is integer-valued).
  
\section{Approximation Analysis}

In this section, we will be considering the execution of the adaptive greedy policy $\sigma$, the optimal policy $\pi^*$, and the hybrid policy $\pi^{\psi}$ on the same random realization $\Phi$.  To make our expressions more compact, we will use the following notation to express, for each of these policies, the event that a subrealization $\psi'$ is visited:

\begin{itemize}
\item $\Gr(\psi')$ denotes the event $\psi' \subset \Psi(\sigma)$ 
\item 
$\Op(\psi')$ denotes the event  $\psi' \subset \Psi(\pi^*)$ 
\item 
 $\Hy(\psi')$ denotes the event $\psi' \subset \Psi(\pi^{\psi})$ 
 \end{itemize}
 
 Consider running the greedy policy and the optimal policy on the same random realization $\Phi$.
 We will use the following notation to express that item $e$ is selected by 
 the optimal policy but not the greedy policy, or that it is selected by both policies:

 \begin{itemize}
     \item $\A(e)$ denotes the event $e \in S(\pi^*) \setminus S(\sigma)$
     \item $\B(e)$ denotes the event $e \in S(\sigma) \cap S(\pi^*)$
 \end{itemize}

The following lemma equates the expected cost of greedy policy $\sigma$ to its expected revenue.
\begin{lemma}
  \label{lemma:dsigma=lambda}
  \begin{gather}
    \E[C(\sigma)] = \sum_{i=1}^{\cardinality{N(\sigma)}} \E[\lambda_i] \label{eqn:dsigma=lambda}
  \end{gather}
\end{lemma}
\begin{proof}
Recall that $e_{\psi} = \sigma(\psi)$.
We have
  
   \begin{align*}
    \sum_{i=1}^{\cardinality{N(\sigma)}} \E[\lambda_i]
    &  {=}~~ \E\left[\sum_{\psi \in N(\sigma) \text{ s.t } \Gr(\psi)} \upsilon_{\psi}
    \sum_{i \text{ s.t. } \psi \leadsto i} \epsilon_i\right] & \text{ by the definition of $\lambda_i$ in \eqref{eqn:lambdai}} \\
    & {=}~~ \E\left[\sum_{\psi \in N(\sigma) \text{ s.t } \Gr(\psi)} \upsilon_{\psi} F_{\psi}(e_{\psi})\right]  & \text{ by Fact~\ref{fact:lambdapsi}}\\
    & = \sum_{\psi \in N(\sigma)} \Pr[\Gr(\psi)]\upsilon_{\psi} \E[F_{\psi}(e_{\psi})) \vert \Gr(\psi)]\\
    & = \sum_{\psi \in N(\sigma)} \Pr[\Gr(\psi)]\upsilon_{\psi} \E[F_{\psi}(e_{\psi})] & \text{ by the Independence Assumption} \\
    & {=} \sum_{\psi \in N(\sigma)} \Pr[\Gr(\psi)]c(e_{\psi}) & \text{ by the definition of } \upsilon_{\psi} \text{ in  \eqref{eqn:greedyrate}}\\
    & = \sum_{e \in E} ~~\sum_{\psi \in \sigma_{-1}[e]} \Pr[\Gr(\psi)]c(e) \\
    & = \sum_{e \in E} \Pr[e \in S(\sigma)] c(e) \\
    & = \E[C(\sigma)]. \qedhere
  \end{align*}
\end{proof}

\bigskip
The following lemma equates the expected cost of $\pi^*$ to its expected revenue.
\begin{lemma}
  \label{lemma:d*>=mu*}
  \begin{gather}
    \E[C(\pi^*)] = \E[\mu^*] 
    \nonumber 
  \end{gather}
\end{lemma}

\begin{proof}
Recalling that $\A(e)$ denotes the event $e \in S(\pi^*) \setminus S(\sigma)$,
define
\begin{gather*}
    J \triangleq \left( \sum_{e \in E} \Pr[\A(e)] c(e) \right).  
\end{gather*}

Recalling that $\B(e)$ denotes the event $e \in S(\pi^*) \cap S(\sigma)$, and using the definitions of $\mu$ and $\mu^*$ in \eqref{eqn:mu} and  \eqref{eqn:mu*},
we have

\bigskip
  \begin{align}
    \E[\mu^*] & {=} \sum_{e \in E} \E[\mu(e)] \nonumber \\
    & {=} 
    J + \Bigg( \sum_{e \in E} \Pr[\B(e)] \sum_{\psi \in \sigma_{-1}[e]} \Pr[\Gr(\psi) \vert \B(e) ] 
    \upsilon_{\psi} \E[F_{\psi}(e) \vert \B(e) \wedge \Gr(\psi)] \Bigg) 
    \label{eqn:J}
    \end{align}
  
  \bigskip  
    To simplify the above expression, we first prove the following:
    
    \begin{align}
    \forall e \in E,~ \psi \in \sigma_{-1}[e]:~ 
    \E[F_{\psi}(e) \vert \B(e) \wedge \Gr(\psi)]=\E[F_{\psi}(e)]
    \label{eqn:claim}
    \end{align}
    \bigskip
    
    To prove~\eqref{eqn:claim},
    assume $e \in E$ and $\psi \in \sigma_{-1}[e]$.  
    Since
    $\psi \in \sigma_{-1}[e]$, 
    $\Gr(\psi)$ implies $e \in S(\sigma)$.  Therefore, the condition $\B(e) \wedge \Gr(\psi)$ can be replaced by
    $e \in S(\pi^*) \wedge \Gr(\psi)$. We can refine this conditioning further by taking into account not just the fact that $e \in S(\pi^*)$, but by considering which subrealization $\psi^*$ in $\pi^*$ satisfies $\psi^* \in \pi^*_{-1}[e] \wedge \Op(\psi^*)$. So:

   \begin{align}
     \E[F_{\psi} & (e) \vert \B(e) \wedge \Gr(\psi)] \nonumber \\
       & =  \E[F_{\psi}(e) \vert e \in S(\pi^*) \wedge \Gr(\psi)] \nonumber \\
    & = \sum_{\psi^* \in \pi^*_{-1}[e]} \Pr[\Op(\psi^*) \vert e \in S(\pi^*) \wedge \Gr(\psi)]\E[F_{\psi}(e) \vert \Op(\psi^*) \wedge \Gr(\psi)]  \nonumber \\
    & 
    {=} 
    \sum_{\psi^* \in \pi^*_{-1}[e]} \Pr[\Op(\psi^*)\vert e \in S(\pi^*) \wedge \Gr(\psi)]
    \E[F_{\psi}(e)]  
    \label{eqn:indepstuff} \\
    & =   
    \E[F_{\psi}(e)]. 
    \label{eqn:partitions} 
  \end{align}
  
  \bigskip
{\noindent Line \eqref{eqn:indepstuff}} holds due to the fact that $\psi \in \sigma_{-1}[e]$ by assumption, and $\psi^* \in \pi^*_{-1}[e]$, so $e$ is not an element of either $dom(\psi)$ or
$dom(\psi^*)$.  
Whether $\sigma$ visits node $\psi$ of $T(\sigma)$ depends only on the states of the items in $dom(\psi)$, 
and whether $\pi^*$ visits node $\psi^*$ of $T(\pi^*)$ depends only on the states of the items in $dom(\psi^*)$.  Thus 
by the Independence Assumption,
the value of $\Phi(e)$, 
and hence the value of $\E[F_{\psi}(e)]$,
is independent of the conditions
$\Gr(\psi)$ and $\Op(\psi^*)$.
Line~\eqref{eqn:partitions} holds because if $\pi^*$ selects $e$ when executed on $\Phi$, it visits exactly one realization
$\psi^* \in \pi^*_{-1}[e]$.
Hence
$\sum_{\psi^* \in \pi^*_{-1}[e]} \Pr[\Op(\psi^*) \vert e \in S(\pi^*) \wedge \Gr(\psi)]
    =1$.  
    This completes the proof of ~\eqref{eqn:claim}.  
 
 \bigskip
 We can now use \eqref{eqn:claim} to rewrite the expression for
     $\E[\mu^*]$ in \eqref{eqn:J}, yielding

  \begin{align}
    \E[\mu^*] & = J + \sum_{e \in E} \Pr[\B(e)] \sum_{\psi \in \sigma_{-1}[e]} \Pr[\Gr(\psi) \vert \B(e) ] 
    \upsilon_{\psi} 
    \E[F_{\psi}(e)]   \nonumber \\
    &
    %
    {=} J
    + \sum_{e \in E} \Pr[\B(e)] \sum_{\psi \in \sigma_{-1}[e]}
    \Pr[\Gr(\psi) \vert \B(e)] c(e) 
    ~~~~~~~~~~~~~~~~~~~~~~~~~~~~~~~~~~~~\text{ by the definition of } \upsilon_{\psi} 
    \nonumber \\
    & = J
    + \Bigg( \sum_{e \in E} \Pr[\B(e)] c(e) \Bigg)  
    ~~~~~~~~~~~~~~~~~~~~~~~~~~~~~~~~~~~~~~~~~~~~
     \text{ because }  \sum_{\psi \in \sigma_{-1}[e]} \Pr[\Gr(\psi) \vert \B(e)] = 1  \nonumber \\
    & = \left( \sum_{e \in E} \Pr[e \in S(\pi^*) \setminus S(\sigma)] c(e) \right) + \Bigg( \sum_{e \in E} \Pr[e \in S(\pi^*) \cap S(\sigma)] c(e) \Bigg)  \nonumber \\
   &  = \sum_{e \in E} \Pr[e \in S(\pi^*)]c(e) \nonumber \\
    & = \E[C(\pi^*)]. \qedhere
    \nonumber 
  \end{align}
\end{proof}

Recall the description of the hybrid policy $\pi^{\psi}$, where $\psi$ is a non-leaf node of $T(\sigma)$.  At node $\psi$, $\sigma$ selects item $e_{\psi}$ and $e_{\psi} \not\in dom(\psi)$.
The tree $T(\pi^{\psi})$ for the hybrid policy also contains node $\psi$, but the item selected at that node may be different from $e_{\psi}$ (since the hybrid policy switches to $\pi^*$ at this point). 

With that background, we are ready to prove the following technical lemma.

\begin{lemma}
  \label{lemma:deltapluslem}
  Let $\psi \in N(\sigma)$ and $i \in \{1, \ldots, \cardinality{N(\sigma)} \}$ such that
  $\Pr[\psi \leadsto i] > 0$. Then
  \begin{gather}
    \forall e \in E \setminus dom(\psi):~
    \alpha\E[\mu(e) \vert \psi \leadsto i]
    \geq
    \E[\delta^+_{\psi}(e) \vert \psi \leadsto i] \upsilon_{\psi} 
  \end{gather}
\end{lemma}

\begin{proof}
Let $e \in E \setminus dom(\psi)$ and let $\psi$ and $i$ be as given in the statement of the lemma.

Define $H \triangleq \{\psi_h  \vert \psi_h \in \pi_{-1}^{\psi}[e] \wedge \psi \subseteq \psi_h\}$.
Thus $H$ is the set of nodes of $T(\pi^{\psi})$ where $e$ could be selected during the second stage of $\pi^{\psi}$.

 We divide our analysis according to whether $e$ is selected by both $\sigma$ and $\pi^*$ when they are executed on $\Phi$, or just by $\pi^*$.  We first prove the inequality in the statement of the lemma with respect to the additional condition 
 $\A(e)$ ($e$ is selected by $\pi^*$ but not by $\sigma$), and then with respect to the additional condition
 $\B(e)$ ($e$ is selected by both $\sigma$ and $\pi^*$).  We do not need to consider conditions where $e$ is not selected by $\pi^*$, because $\mu(e)$ and $\delta_{\psi}^+(e)$ are equal to 0 in these cases.
 
 \bigskip
 \begin{itemize}
\item Additional condition $\A(e)$

We will show
\begin{align}
\label{eqn:A(e)}
\alpha\E[\mu(e) \vert \A(e) \wedge \psi \leadsto i] \geq \upsilon_{\psi} \E[\delta_{\psi}^+(e) \vert \A(e) \wedge \psi \leadsto i]
\end{align}
 
 Recall that $\A(e)$ denotes the event that $e$ is in $S(\pi^*)$ but not in $S(\sigma)$. 
 The event $\psi \leadsto i$ implies that $\psi \subset \Psi(\sigma)$.  
 Thus 
 by Fact~\ref{fact:opttohybrid}, 
 $\A(e) \wedge \psi \leadsto i$
 implies that
 $e \in S(\pi^{\psi})$.  Further, since $e \not\in dom(\psi)$, $e$ must be selected in the second stage of $\pi^{\psi}$.
 
  Let $X \subseteq L(\sigma)$ denote the set of terminal subrealizations  observable (covers constructed) by $\sigma$
  given that $\A(e) \wedge  \psi \leadsto i$.  Thus $X$ is a subset of the possible values of $\Psi(\sigma)$. We have
  
  \begin{align*}
    \alpha\E[& \mu(e) \vert \A(e) \wedge \psi \leadsto i] \\
    & = \sum_{\psi_h \in \pi^{\psi}_{-1}[e]} ~~\sum_{\psi_x \in X}
    \Pr[\Hy(\psi_h) \wedge \Psi(\sigma) = \psi_x \vert  \A(e) \wedge \psi \leadsto i ] \alpha c(e) \nonumber & 
    \\
    & ~~~~~~~\text{ since } \mu(e)=c(e) \text{ when } \A(e) \text{ holds, and partitioning event } (\A(e) \wedge \psi \leadsto i) \text{ w.r.t. which } \ 
    \\ & ~~~~~~~~~~~~~\psi_h \in \pi_{-1}^{\psi}[e] 
    \text{ is visited by } \pi^{\psi} \text{ and which cover } \psi_x \in X \text{ is constructed by } \sigma \\
    & 
    {\geq} \sum_{\psi_h \in \pi^{\psi}_{-1}[e]} ~~\sum_{\psi_x \in X}
    \Pr[\Hy(\psi_h) \wedge \Psi(\sigma) =  \psi_x \vert \A(e) \wedge \psi \leadsto i ] \upsilon_{\psi} \E[F_{\psi}(e)] \nonumber ~~~\text{ by the greedy choice \eqref{eqn:greedyrate}}
    \\
    & 
    {=} \sum_{\psi_h \in \pi^{\psi}_{-1}[e]} ~~\sum_{\psi_x \in X}
    \Pr[\Hy(\psi_h) \wedge \Psi(\sigma) = \psi_x \vert \A(e) \wedge \psi \leadsto i ] \upsilon_{\psi} \E[F_{\psi}(e) \vert \Hy(\psi_h) \wedge \Psi(\sigma)=\psi_x] \nonumber & \\
    & ~~~~~~~~\text{ by the Independence Assumption since } e \not\in dom(\psi_h) \cup dom(\psi_x)\\
    & 
    {\geq} \sum_{\psi_h \in \pi^{\psi}_{-1}[e]} ~~\sum_{\psi_x \in X}
    \Pr[\Hy(\psi_h) \wedge \Psi(\sigma)=\psi_x \vert \A(e) \wedge \psi \leadsto i ] \upsilon_{\psi} \E[F_{\psi_h}(e) \vert \Hy(\psi_h) \wedge \Psi(\sigma) = \psi_x] \nonumber \\
    & ~~~~~~~~\text{ replacing } F_{\psi} \text{ by } F_{\psi_h}, \text{ by submodularity since } \psi \subseteq \psi_h \\
    & =  \upsilon_{\psi} \sum_{\psi_h \in \pi^{\psi}_{-1}[e]} ~~\sum_{\psi_x \in X}
    \Pr[\Hy(\psi_h) \wedge \Psi(\sigma)=\psi_x \vert \A(e) \wedge \psi \leadsto i ]  \E[\delta_{\psi}^+(e) \vert \Hy(\psi_h) \wedge \Psi(\sigma) = \psi_x] \nonumber \\
    & ~~~~~~~~\text{ because } \Hy(\psi_h) \text{ implies } e \text{ was selected by } \pi^{\psi} \text{ at node } \psi_h \text{ during its second stage } \\
    & =  \upsilon_{\psi} 
    \E[\delta^+_{\psi}(e)  \vert \A(e) \wedge \psi \leadsto i ]. 
  \end{align*}

\bigskip
\item Additional condition $\B(e)$ 

Because $e \not\in dom(\psi)$, condition $\B(e) \wedge \psi \leadsto e$ implies, by Fact~\ref{fact:opttohybrid},  that $\pi^{\psi}$  visits some $\psi_h \in H$.

We will show
\begin{align}
\label{eqn:B(e)}
\alpha\E[\mu(e) \vert \B(e) \wedge \psi \leadsto i] \geq \upsilon_{\psi} \E[\delta_{\psi}^+(e) \vert \B(e) \wedge \psi \leadsto i]
\end{align}

We first give the proof of \eqref{eqn:B(e)} for the case where $e = e_{\psi}$.  If $e=e_{\psi}$, we have
\begin{align*} 
 & \E[\mu(e_{\psi}) \vert \psi \leadsto i] \\
  &  =  \Pr[e_{\psi} \in S(\pi^*) \vert \psi \leadsto i] \E[\mu(e_{\psi}) \vert \psi \leadsto i \wedge e_{\psi} \in S(\pi^*) ] && \text{by the definition of } \mu 
  \nonumber \\
  & {=} 
    \Pr[e_{\psi} \in S(\pi^{\psi}) \vert \psi \leadsto i] \E[\mu(e_{\psi}) \vert \psi \leadsto i \wedge e_{\psi} \in S(\pi^{\psi}) ] &&\text{ by Fact ~\ref{fact:opttohybrid}}\nonumber \\
 & =  \sum_{\psi_h \in H}
    \Pr[\Hy(\psi_h) \vert \psi \leadsto i ]
    \E[\mu(e_{\psi}) \vert \psi \leadsto i \wedge \Hy(\psi_h)] & &
    \text{ partitioning
    w.r.t. which } \psi_h \in H \text{ is visited}\\
    & {=} \sum_{\psi_h \in H} \Pr[\Hy(\psi_h) \vert \psi \leadsto i ]
    \upsilon_{\psi} \E[F_{\psi}(e_{\psi}) \vert \psi \leadsto i \wedge \Hy(\psi_h)] && \text{ by the definition of } \mu \text{ since } \psi \leadsto i \Rightarrow \Gr(\psi)
    \nonumber \\
    & {\geq}  \sum_{\psi_h \in H} \Pr[\Hy(\psi_h) \vert \psi \leadsto i ]
    \upsilon_{\psi} \E[F_{\psi_h}(e_{\psi})  \vert \psi \leadsto i \wedge \Hy(\psi_h)] \nonumber 
    && \text{ replacing } f_{\psi} \text{ by } f_{\psi_h}, \text{ by submodularity} \\ 
   & = \upsilon_{\psi}  \E[\delta^+_{\psi}(e_{\psi}) \vert \psi \leadsto i] && \text{ by  definition of } \delta^+_{\psi}. 
  \end{align*}
Thus~\eqref{eqn:B(e)} holds when $e = e_{\psi}$.

So suppose $e \neq e_{\psi}$. In this case,
define $G$ to be the set of subrealizations $\psi_g$ satisfying the following three properties:
\begin{gather*}
   \psi_g \in \sigma_{-1}[e] \\
    \psi_g \supset \psi  \\
    f(\psi \cup (e_{\psi}, \psi_g(e_{\psi}))) > f(\rho_i)
   \end{gather*}
   The second property says that in the greedy tree, node $\psi$ is an ancestor of node $\psi_g$. The third property concerns
   the child of node $\psi$ that is on the path from $\psi$ down to $\psi_g$: this is the node identified with subrealization
   $\psi \cup (e_{\psi}, \psi_g(e_{\psi}))$.  It states that the value of $f$ on this subrealization is strictly greater than $f(\rho_i)$.  Therefore,
  \begin{align}
      \forall \psi_g \in G:~ \Gr(\psi_g) \Rightarrow \psi \leadsto i
      \label{eqn:gtopsii}
  \end{align}
  Conversely, if $\psi \leadsto i$, then because $e \not\in dom(\psi)$, condition $\B(e)$ implies that $\sigma$ must visit some $\psi_g \in G$. As before, condition $\B(e) \wedge \psi \leadsto i$ implies that $\pi^{\psi}$ must visit some $\psi_h \in H$.
  Therefore, 
  \begin{align}
      \B(e) \wedge \psi \leadsto i \Rightarrow \exists! \psi_g \in G, ~\exists! \psi_h \in H \text{ such that } \Gr(\psi_g) \wedge \Hy(\psi_h) \nonumber
  \end{align}
 where $\exists!$ means the quantified element exists and is unique.  
 
 Note that by the Independence Assumption, for $\psi_g \in G$ and $\psi_h \in H$, since $e \not\in dom(\psi_g) \cup dom(\psi_h)$,
 \begin{align}
 \label{eqn:Fg}
 \E[F_{\psi_g}(e) \vert \Gr(\psi_g) \wedge \Hy(\psi_h)] = \E[F_{\psi_g}(e)]   
 \end{align}
 
Before proceeding with the proof of \eqref{eqn:B(e)},
we show that the following equation holds for 
 $\psi_g \in G:$
 \begin{align}
 \label{eqn:necessary}
 \E[\mu(e) \vert \B(e) \wedge \Gr(\psi_g)] = c(e) 
 \end{align}

 The proof of \eqref{eqn:necessary} is as follows.  Recalling
that $e \not\in dom(\psi)$, 
\begin{align}
& \E[\mu(e) \vert \B(e) \wedge \Gr(\psi_g)] 
\nonumber \\ 
&= \upsilon_{\psi_g} \E[F_{\psi_g}(e) \vert \B(e) \wedge \Gr(\psi_g)] ~~~~\text{ by the definition of $\mu(e)$, since $\psi_g \in \sigma_{-1}[e]$ }\nonumber
\\ &=  \sum_{\psi_h \in H} P[\Hy(\psi_h) \vert \B(e) \wedge \Gr(\psi_g)] \upsilon_{\psi_g} \E[F_{\psi_g}(e) \vert \Gr(\psi_g) \wedge \Hy(\psi_h)] \nonumber
\\& ~~~~~~~~~~~~\text{ partitioning w.r.t. which } \psi_h \in H \text{ is visited by } \pi^{\psi}\nonumber
\\ &=  \sum_{\psi_h \in H} P[\Hy(\psi_h) \vert \B(e) \wedge \Gr(\psi_g)] \upsilon_{\psi_g} \E[F_{\psi_g}(e_{\psi_g})] \nonumber
\\ &~~~~~~~~~~~~~~~~~~~\text{ by } \eqref{eqn:Fg} \text{ and because } \psi_g \in \sigma_{-1}[e], \text{ so } e = e_{\psi_g} \nonumber
\\ &{=}  c(e) \nonumber
\\ & ~~~~~~~~~~~~\text{ because } \sum_{\psi_h \in H} P[\Hy(\psi_h) \vert \B(e) \wedge \Gr(\psi_g)] = 1 \text{ and } \upsilon_{\psi_g} = c(e_{\psi_g})/\E[F_{\psi_g}(e_{\psi_g})].
\nonumber
\end{align}

\bigskip
Having proved \eqref{eqn:necessary},
we now prove \eqref{eqn:B(e)}, the desired inequality for condition $\B(e)$.  For $\psi_g \in G$, define
\begin{gather*}
    K_g \triangleq P[\Gr(\psi_g) \vert \B(e) \wedge \psi \leadsto i]
\end{gather*}
We have
\begin{align*}
&\alpha\E[\mu(e) \vert \B(e) \wedge \psi \leadsto i] & \nonumber \\
& =  \alpha \sum_{\psi_g \in G} P[\Gr(\psi_g) \vert \B(e) \wedge \psi \leadsto i] \E[\mu(e) \vert \B(e) \wedge \Gr(\psi_g)] 
\\& ~~~~~~~~~\text{ partitioning w.r.t. which } \psi_g \in G \text{ is visited, and because } \Gr(\psi_g) \Rightarrow \psi \leadsto i\nonumber
\\ &{=}  \alpha \sum_{\psi_g \in G} K_g~c(e) \nonumber~~~~~~~~\text{ by \eqref{eqn:necessary} and the definition of } K_g
\\ &
{=}  \alpha \sum_{\psi_g \in G} K_g \sum_{\psi_h \in H} P[\Hy(\psi_h) \vert \B(e) \wedge \Gr(\psi_g)]\upsilon(e \vert\psi_h) \E[F_{\psi_h}(e)] \nonumber
\\& ~~~~~~~~\text{ partitioning w,r.t. which } \psi_h \in H \text{ is visited, and because } \upsilon(e \vert \psi_h) = c(e) / \E[F_{\psi_h}(e)]
 \\
 & = \alpha \sum_{\psi_g \in G} K_g \sum_{\psi_h \in H} P[\Hy(\psi_h) \vert \B(e) \wedge \Gr(\psi_g)]\upsilon(e \vert \psi_h)\E[F_{\psi_h}(e) \vert \Gr(\psi_g)  \wedge \Hy(\psi_h)] \nonumber
 \\& ~~~~~~~~~~\text{ by the Independence Assumption } \\
 & = \alpha \sum_{\psi_g \in G} K_g \sum_{\psi_h \in H} P[\Hy(\psi_h) \vert \B(e) \wedge \Gr(\psi_g)]\upsilon(e \vert \psi_h)\E[\delta_{\psi}^+(e) \vert \Gr(\psi_g)  \wedge \Hy(\psi_h)] \nonumber
 \\
 &~~~~~~~~~~\text{ by the definition of } \delta^+_{\psi}(e), \text{ since  } \Hy(\psi_h) \text{ implies } e \text{  was selected by }
 \\& ~~~~~~~~~~~~~~~~~~~~~~~~~~~~~ \pi^{\psi} \text{ at node } \psi_h \text{ during its  second stage } 
 \\
 & {\geq}  \alpha \upsilon(e \vert \psi) \sum_{\psi_g \in G} K_g \sum_{\psi_h \in H} P[\Hy(\psi_h) \vert \B(e) \wedge \Gr(\psi_g)]\E[\delta_{\psi}^+(e) \vert \Gr(\psi_g) \wedge \Hy(\psi_h)] 
 \\& ~~~~~~~~~
 \text{ since } \psi \subseteq \psi_h \text{ so by the submodularity property, } F_{\psi}(e) \geq F_{\psi_h}(e)
 \\
 &~~~~~~~~~~~~~~~~~~~~~~~~~~~~~~~~~~~~~~ \text{ and hence } \upsilon(e \vert \psi_h) \geq \upsilon(e \vert \psi)
\\
&{\geq}   \upsilon_{\psi} \sum_{\psi_g \in G} K_g \sum_{\psi_h \in H} P[\Hy(\psi_h) \vert \B(e) \wedge \Gr(\psi_g)]\E[\delta_{\psi}^+(e) \vert \Gr(\psi_g) \wedge \Hy(\psi_h)]
\\& ~~~~~~~~~~\text{ since }  \upsilon_{\psi} = \upsilon(e_{\psi} \vert \psi) \leq \alpha \upsilon(e \vert \psi) \text{ by the greedy choice } \eqref{eqn:greedyrate}  
\\
&  = \upsilon_{\psi} \sum_{\psi_g \in G} K_g \E[\delta_{\psi}^+(e) \vert \B(e) \wedge \Gr(\psi_g)] \nonumber \\
 &=   \upsilon_{\psi} \sum_{\psi_g \in G} P[\Gr(\psi_g) \vert \B(e) \wedge \psi \leadsto i] \E[\delta_{\psi}^+(e) \vert \B(e) \wedge \Gr(\psi_g)]
\\& ~~~~~~~~~~\text{ by the definition of } K_g 
 \\
& =  \upsilon_{\psi} \E[\delta_{\psi}^+(e) \vert \B(e) \wedge \psi \leadsto i]. \nonumber
\end{align*}

  \end{itemize}
Finally, because $\mu(e)=0$ and $\delta_{\psi}^+(e)=0$ whenever $e \not\in S(\pi^*)$, the lemma follows from  the inequalities $\eqref{eqn:B(e)}$ and $\eqref{eqn:A(e)}$ proved for the two conditions, $\A(e)$ and $\B(e)$.
\end{proof}

\bigskip
\begin{fact}
  \label{fact:atleastmin}
  For non-negative numbers $\alpha_1, \ldots, \alpha_n$, and $\beta_1, \ldots, \beta_n$ such that
  $\sum_{i = 1}^{n} \beta_i > 0$:
  \begin{gather}
    \min_{i \text{ s.t. } \beta_i > 0} ~~\frac{\alpha_i}{\beta_i}
    \leq \frac{\sum_{i=1}^{n} \alpha_i}{\sum_{i=1}^{n} \beta_i} \label{eqn:atleastmin}
  \end{gather}
\end{fact}

We combine the above fact with Lemma~\ref{lemma:deltapluslem} to prove the following lemma.

\bigskip
\begin{lemma}
  \label{lemma:conditionallambdai}
  \begin{gather}
    \forall i \in \{1, \ldots, \cardinality{N(\sigma)} \}:~\E[\lambda_i \vert \psi \leadsto i]
    \leq \frac{\alpha\E[\mu^* \vert \psi \leadsto i]}{Q - f(\rho_i)}\epsilon_i 
  \end{gather}
\end{lemma}
\begin{proof}
  Let $E_1 = \{e \vert e \in E\setminus dom(\psi) \wedge \E[\delta^+_{\psi}(e) \vert \psi \leadsto i] > 0\}$.
  \begin{align*}
    \E[\lambda_i \vert \psi \leadsto i]
     = \upsilon_{\psi}\epsilon_i
   &  
   {\leq} \min_{e \in E_1} \frac{\alpha\E[\mu(e) \vert \psi \leadsto i]}{\E[\delta^+_{\psi}(e) \vert \psi \leadsto i]} \epsilon_i  ~~~~~~~~~~~ \text{ by Lemma }  \ref{lemma:deltapluslem} \\
   & 
    {\leq} \frac{\sum_{e \in E \setminus dom(\psi)} \alpha\E[\mu(e) \vert \psi \leadsto i]}{\sum_{e \in E \setminus dom(\psi)} \E[\delta^+_{\psi}(e) \vert \psi \leadsto i]} \epsilon_i ~~~~~~~~~~~~~ \text{ by Fact } \ref{fact:atleastmin} \\
    & 
    {\leq} \frac{\alpha\E[\mu^* \vert \psi \leadsto i]}{Q - f(\psi)} \epsilon_i~~~~~~~~~\text{ by  definitions of } \mu^* \text{ and } \delta_{\psi}^+  \qedhere
  \end{align*}
\end{proof}

\begin{lemma}
  \label{lemma:unconditionallambdai}
  \begin{gather}
    \forall i \in \{1, \ldots, \cardinality{N(\sigma)} \}:~\E[\lambda_i] \leq \alpha \E[C(\pi^*)] \frac{\epsilon_i}{Q - f(\rho_i)} 
  \end{gather}
\end{lemma}
\begin{proof}
Fix $i \in \{1, \ldots,  |N(\sigma)|\}$.
   Let $R = \{\psi \vert \psi \in N(\sigma) \wedge \Pr[\psi \leadsto i] > 0\}$.
   The events $\{\psi \leadsto i \vert \psi \in R \}$ partition the sample space. Hence:
  \begin{align*}
    \E[\lambda_i]
    & = \sum_{\psi \in R} \Pr[\psi \leadsto i] \E[\lambda_i \vert \psi \leadsto i] \\
    & 
    {\leq}
    \sum_{\psi \in R} \frac{\Pr[\psi \leadsto i] \alpha \E[\mu^* \vert \psi \leadsto i] \epsilon_i}{Q - f(\rho_i)} && \text{ by Lemma }\ref{lemma:conditionallambdai}\\
    &= \frac{\alpha \E[\mu^*] \epsilon_i}{Q - f(\rho_i)} \\
    & =
    \frac{\alpha \E[C(\pi^*)]\epsilon_i}{Q - f(\rho_i)} && \text{ by Lemma } \ref{lemma:d*>=mu*} \qedhere
  \end{align*}
\end{proof}

\begin{lemma}
  \label{lemma:epsilon}
  \begin{gather}
    \E[C(\sigma)] \leq \alpha\E[C(\pi^*)] \sum_{i = 1}^{\cardinality{N(\sigma)}} \frac{\epsilon_i}{Q - f(\rho_i)} 
  \end{gather}
\end{lemma}
\begin{proof}
  The lemma follows by combining Lemmas \ref{lemma:dsigma=lambda} and \ref{lemma:unconditionallambdai}.
\end{proof}

With the above lemmas, we can now prove Theorem~\ref{theo:greedy},
which states that $\sigma$ achieves an approximation bound of
$\alpha\kappa$ for Stochastic Submodular Cover, where $\kappa = \ln (Q/\eta)+1$
when the utility function is real valued and $\kappa = H(Q)$ when the utility function is integer valued.

\bigskip
{\bf \noindent Proof of Theorem~\ref{theo:greedy}:}
 The theorem follows from Lemma~\ref{lemma:epsilon} by bounding the summation
 
 \begin{gather*}
 \sum_{i = 1}^{\cardinality{N(\sigma)}} \frac{\epsilon_i}{Q - f(\rho_i)}
 \end{gather*}
 using an argument of Azar and Gamzu~\citeyear{azargamzu11}, which we include for completeness.

  Let $t$ denote $|N(\sigma)|$.  Recall that $\epsilon_i = f(\rho_{i+1}) - f(\rho_i)$.
For $i \in \{1, \ldots, t-1\}$,
  
  $$\frac{\epsilon_i}{Q-f(\rho_i)} = \int_{f(\rho_i)}^{f(\rho_{i+1})} \frac{1}{Q-f(\rho_i)} dx \leq  \int_{f(\rho_i)}^{f(\rho_{i+1})} \frac{1}{Q-x} dx.$$

  Therefore,
  \begin{align}
  \sum_{i = 1}^{t} \frac{\epsilon_i}{Q - f(\rho_i)} \nonumber & \leq 
 \frac{\epsilon_t}{Q-f(\rho_t)} + \int_{0}^{f(\rho_t)} \frac{1}{Q-x} dx \nonumber \\
  &= 1 + \ln~Q/\epsilon_t~~~\text{ because } \epsilon_t = Q - f(\rho_t)  
  \nonumber \\
  & \label{eqn:fracbound}
  \leq 1 + \ln~Q/\eta~~~\text{ by } (\ref{eqn:lastepsilon})
  \end{align}
  
  \smallskip
  If $f$ is integer-valued, then
  $\epsilon_t \geq 1$ and
  similarly,
  
  \begin{align}
  \sum_{i = 1}^{t} \frac{\epsilon_i}{Q - f(\rho_i)} &\leq 1 + \sum_{j=0}^{f(\rho_t)-1} \frac{1}{Q-j} \nonumber \\
  &= 1 + H(Q) - H(\epsilon_t) \nonumber \\
  \label{eqn:intbound}
  & \leq H(Q).
  \end{align}

\qed

\section*{Acknowledgments}
Partial support for this work came from 
NSF Award IIS-1909335 (for L. Hellerstein) and from a PSC-CUNY Award, jointly funded by The Professional Staff Congress and The City University of New York (for D. Kletenik).
We thank the anonymous referees for their comprehensive and useful comments.  We thank Naifeng Liu for proofreading help.

\appendix

\section{The Definition of the Stochastic Submodular Cover Problem}
\label{app:def}

In the definition of Stochastic Submodular Cover given by 
Golovin and Krause (who called the problem Stochastic Submodular {\em Coverage}), the input utility function 
is $\hat{f}:2^{E \times O} \rightarrow \mathbb{R}_{\geq 0}$
where $E$ is the set of items and $O$ is the set of states,
such that $\hat{f}$ is a monotone and submodular set function.
Since $\hat{f}$ is defined on all subsets of $E \times O$, 
its domain includes ``inconsistent'' subsets containing pairs $(e,o)$
and $(e,o')$, where $o \neq o'$.
Intuitively, this means that the input utility function must assign values to collections of items in which a single item can simultaneously be in more than one state; further,  these values must be compatible with the properties of monotonicity and submodularity.
In contrast, our input utility function need only assign values to collections of items where each item has exactly one state.  In fact, though,
the value of $\hat{f}$ on the ``inconsistent'' subsets is irrelevant both to the remainder of 
the Golovin and Krause problem definition, 
and to the proof of their bound for Stochastic Submodular Cover.  While in their definition they require $\hat{f}$ to have values on these subsets, they effectively ignore these values. 

The definition of Stochastic Submodular Cover given by Golovin and Krause,
with item set $E = \{e_1, \ldots, e_n\}$ and state set $O$, is equivalent to a special case of the Weighted Stochastic Submodular Ranking (WSSR) problem, 
studied by Im et al.~\citeyear{imetal}.  In particular, it is equivalent to the special case of the WSSR problem 
where there is a single function $f_1:2^D \rightarrow [0,1]$ with $D = E \times O$, and such that for each $X_i$, $P[X_i = (e,o)]= 0$ if $e \neq e_i$, otherwise $P[X_i = (e,o)] $ equals the probability $P[\Phi(e_i) = o]$ in Stochastic Submodular Cover.  
Although the problem definition for the WSSR problem requires that $f_1$ be defined on all subsets of $D = E \times O$ in this case, including the ``incompatible'' subsets, the analysis of Im et al.\ applied to this case makes no use of the value of $f_1$ on such subsets.  

\section{The Error in the Analysis of \shortciteR{Liu:2008:NAS:1376616.1376633}}
\label{app:error}

The paper of Liu et al.\ claims to show an $(\ln~m + 1)$ approximation bound for the adaptive greedy policy, applied to Stochastic Set Cover.
Consider the execution of this policy, which always selects the stochastic subset covering the largest expected number of items, per unit cost.  Let $S_{i_1}, \ldots, S_{i_k}$ denote the sequence of chosen subsets.  Each $S_{i_\ell}$ covers
some number $h$ of ground elements in $S_{i_\ell}$ not already covered by $S_{i_1}, \ldots, S_{i_{\ell-1}}$.
Consider the $h$ elements to be covered one by one by $S_{i_\ell}$.
The {\em price} paid to cover each one is $\frac{c_{i_{\ell}}}{h}$.

Consider a run of the greedy policy, and of an optimal policy OPT, on a random assignment of values to the $S_i$.  Order the elements of the ground set in the order they are covered by the greedy policy.  
Lemma 3 of Liu et al.\ gives an upper bound on $price(j)$, the price paid when the $j$th item in this order is covered.  The bound is conditioned on $\psi$ and $\Upsilon$, where $\psi$ is the sequence of $S_i$ chosen by the greedy policy (together with the observed values of those $S_i$), and $\Upsilon$ is the subset of $S_i$ chosen by both the greedy policy and OPT.

\begin{figure}[hbt]
\centering
\includegraphics[scale=0.3]{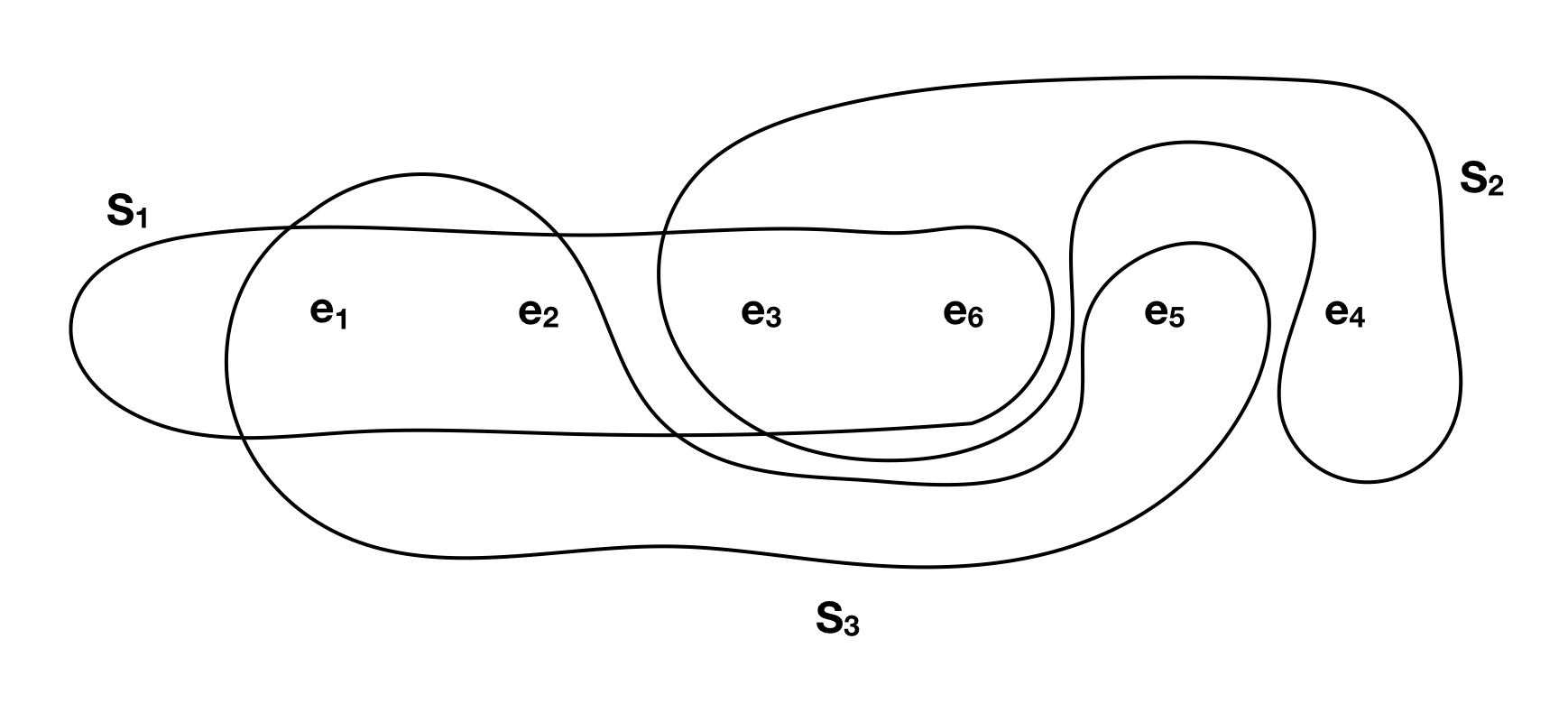}
\caption{Greedy cover example}
\label{fig:counterfig}
\end{figure}

For example,
suppose the set of ground elements is $\{e_1, \ldots, e_6\}$ and there are only three stochastic subsets, $S_1, S_2, S_3$. 
Consider the following realization of these subsets: $S_1 = \{e_1, e_2, e_3, e_6\}$, $S_2 = \{e_3, e_4, e_6\}$, and $S_3 = \{e_1, e_2, e_5\}$.  Suppose that under this realization, the greedy policy covers the ground elements using all three subsets, in the following order: $S_1, S_3, S_2$.  Suppose that the optimal policy, under the above realization, covers the elements using just $S_2$ and $S_3$.  
Figure~\ref{fig:counterfig} shows the $m=6$ elements of the ground set from left to right in the order they were covered by the $S_i$ in the greedy policy (elements covered during the same step are listed in order of their indices).  Here $\psi$ is equal to $S_1, S_3, S_2$ with the realizations listed above, and
$\Upsilon = \{S_2, S_3\}$.

Right before the greedy policy covers its $j$th element, there remain
$m-j+1$ uncovered elements.  Fix $j$ and let $M_{last}$ denote that set of elements.  Let $M_{com} \subseteq M_{last}$ denote the subset of elements in $M_{last}$ that are first covered, during execution of the greedy policy, by subsets in $\Upsilon$.  
A crucial step of the proof of Lemma 3 is to show that

$$m-j+1 = \sum_{S_i \in \Upsilon} cov(S_i) + \sum_{S_i \in \mathcal{S} \backslash \psi} optcov(S_i)$$

{\noindent where} $cov(S_i)$ is the number of elements in $M_{last}$ that are first covered by $S_i$ during execution of the greedy policy, and $optcov(S_i)$ is
the expected number of elements in $M_{last} \backslash M_{com}$ first covered by $S_i$ during execution of the optimal algorithm.
However, this equation does not hold.  (It would be sufficient to show
that the left hand side of the equation is bounded above by the right hand side, but that does not hold either.)

Figure~\ref{fig:counterfig} with $j=1$ is a counterexample.  In this case, $M_{last}$ consists of all 6 ground elements.
In this $\psi$, $S_1$ covers 4 new elements, $S_3$ covers 1 (namely $e_5$), and $S_2$ covers 1 (namely $e_4)$.   
Therefore, $cov(S_2) = 1$ and $cov(S_3) = 1$.  However, $\mathcal{S} \backslash \psi$ is empty.  So in this case, the left hand side of the equation equals 6, while the right hand side equals 2.  
The problem is that in the optimal policy, $S_2$ and $S_3$ are used to cover all 6 ground elements, but the greedy policy
uses them to cover only 2 elements, because the others were already covered by an $S_i$ not in $\Upsilon$.
The conditioning on $\psi$ and $\Upsilon$ does not imply the equality.
While the condition can be strengthened, it is not clear how to do so in a way that doesn't invalidate other
parts of the proof.

\newpage
\section{Table of Frequently Used Symbols}
\label{sec:symbols}
\begin{center}
   \begin{table}[hb]
   \begin{tabular}{c|p{3.00in}}
   Symbol & Description  \\ \hline
   $E, e$ & Set of items, an individual item \\
 $c(e)$ & Cost of item $e$ \\
   $O$ & Set of possible states for an item \\
   $\Phi$ & Random realization of all item states\\
   $\varphi$ & A fixed realization; a possible outcome for $\Phi$ \\
   $\psi$ & A fixed subrealization \\
   $dom(\psi)$ & Items whose states are recorded in $\psi$ \\
   $f$ & Utility function mapping from set of all subrealizations to  $\mathbb{R}_{\geq 0}$ \\
   $F_{\psi}(e)$ & Random marginal increase in $f$ from observing state of $e$, after observing states in $\psi$ \\ 
   $Q$ & Value of $f(\varphi)$ for all realizations $\varphi$ (cover value)
    \\ 
    $\eta$ & Minimum value of $Q-f(\psi)$ for any $f(\psi) < Q$ \\
    $\pi, \pi^*, \sigma$ & A fixed policy; optimal policy;  greedy policy \\
    $\pi^{\psi}$ & Hybrid policy, runs $\sigma$ and switches to $\pi^*$ at $\psi$ \\
   $\pi(\psi)$ & Item selected by $\pi$ after observing states in $\psi$ 
   \\
   $T(\pi)$ & Decision tree for policy $\pi$ \\
   $L(\pi)$, $N(\pi)$ &  Sets of subrealizations for leaf and non-leaf nodes of $T(\pi)$, 
   \\
   $\Psi(\pi)$& Random cover constructed by $\pi$ \\
   $C(\pi)$ & Random cost of policy $\pi$
   \\
   $\mu(e)$, $\mu^*$ & Revenue from $e$ in $\pi^*$; total revenue in $\pi^*$
   \\
   $\lambda_i$ & Revenue from marker $\rho_i$ in $\sigma$
   \\
   $e_{\psi}$ & Item chosen at node $\psi$ of greedy tree $T(\sigma)$
   \\
   $\upsilon(e \vert \psi)$, $\upsilon_{\psi}$ & unit price $c(e)/F_{\psi}(e)$ of $e$ w.r.t. $\psi$; unit price paid at node $\psi$ of $T(\sigma)$
   \\
   $\rho_i$, $\epsilon_i$ & Subrealization that is $i$th marker of $\sigma$; difference in utility between $\rho_i$ and $\rho_{i+1}$
   \\
   $\delta^+_{\psi}(e)$ & Equal to marginal utility from $e$ in hybrid policy, if $e$ selected during second stage, else equals 0 \\
   $\pi_{-1}[e]$ & Subrealizations for nodes of greedy tree where $e$ is chosen \\
   $\psi \leadsto i$ & $\psi$ is last realization visited by $\sigma$ such that $f(\psi) \leq f(\rho_i)$ \\
   $\Gr(e), \Hy(e), \Op(e)$ & $e$ is chosen by: policy $\sigma$,  policy $\pi^{\psi}$, policy $\pi^*$ \\
   $\A(e),\B(e)$ & $e$ is chosen by $\pi^*$ but not by $\sigma$, $e$ is chosen by both
   \end{tabular}
   \caption{Frequently used symbols}
   \end{table}
 \end{center}
 
 \bibliography{greedy}

\begin{thebibliography}{}

\bibitem[\protect\BCAY{Azar\ \BBA\ Gamzu}{Azar\ \BBA\
  Gamzu}{2011}]{azargamzu11}
Azar, Y.\BBACOMMA\  \BBA\ Gamzu, I. \BBOP2011\BBCP.
\newblock \BBOQ Ranking with submodular valuations\BBCQ\
\newblock In {\Bem Proceedings of the {ACM-SIAM} Symposium on Discrete
  Algorithms, {SODA}}, \BPGS\ 1070--1079. {SIAM}.

\bibitem[\protect\BCAY{Deshpande, Hellerstein,\ \BBA\ Kletenik}{Deshpande
  et~al.}{2016}]{Deshpande:2016:AAS:2930058.2876506}
Deshpande, A., Hellerstein, L., \BBA\ Kletenik, D. \BBOP2016\BBCP.
\newblock \BBOQ Approximation algorithms for stochastic submodular set cover
  with applications to {B}oolean function evaluation and min-knapsack\BBCQ\
\newblock {\Bem ACM Trans. Algorithms}, {\Bem 12\/}(3), 42:1--42:28.

\bibitem[\protect\BCAY{Dinur\ \BBA\ Steurer}{Dinur\ \BBA\
  Steurer}{2014}]{DinurSteurer}
Dinur, I.\BBACOMMA\  \BBA\ Steurer, D. \BBOP2014\BBCP.
\newblock \BBOQ Analytical approach to parallel repetition\BBCQ\
\newblock In {\Bem Proceedings of the ACM Symposium on Theory of Computing,
  {STOC}}, \BPGS\ 624--633.

\bibitem[\protect\BCAY{Esfandiari, Karbasi,\ \BBA\ Mirrokni}{Esfandiari
  et~al.}{2019}]{Esfandiarietal19}
Esfandiari, H., Karbasi, A., \BBA\ Mirrokni, V.~S. \BBOP2019\BBCP.
\newblock \BBOQ Adaptivity in adaptive submodularity\BBCQ\
\newblock {\Bem CoRR}, {\Bem abs/1911.03620}.

\bibitem[\protect\BCAY{Feige}{Feige}{1998}]{Feige98}
Feige, U. \BBOP1998\BBCP.
\newblock \BBOQ A threshold of ln \emph{n} for approximating set cover\BBCQ\
\newblock {\Bem J. {ACM}}, {\Bem 45\/}(4), 634--652.

\bibitem[\protect\BCAY{Fujito}{Fujito}{2000}]{fujitoSurvey}
Fujito, T. \BBOP2000\BBCP.
\newblock \BBOQ Approximation algorithms for submodular set cover with
  applications\BBCQ\
\newblock {\Bem IEICE Transactions on Informations and Systems}, {\Bem 83}.

\bibitem[\protect\BCAY{Goemans\ \BBA\ Vondr{\'{a}}k}{Goemans\ \BBA\
  Vondr{\'{a}}k}{2006}]{goemans:stochasticcovering}
Goemans, M.\BBACOMMA\  \BBA\ Vondr{\'{a}}k, J. \BBOP2006\BBCP.
\newblock \BBOQ Stochastic covering and adaptivity\BBCQ\
\newblock In {\Bem Proceedings of LATIN 2006:Theoretical Informatics}, \BPGS\
  532--543. Springer Berlin Heidelberg.

\bibitem[\protect\BCAY{Golovin\ \BBA\ Krause}{Golovin\ \BBA\
  Krause}{2011}]{Golovin:2011:AST:2208436.2208448}
Golovin, D.\BBACOMMA\  \BBA\ Krause, A. \BBOP2011\BBCP.
\newblock \BBOQ Adaptive submodularity: Theory and applications in active
  learning and stochastic optimization\BBCQ\
\newblock {\Bem J. Artif. Int. Res.}, {\Bem 42\/}(1), 427--486.

\bibitem[\protect\BCAY{Golovin\ \BBA\ Krause}{Golovin\ \BBA\
  Krause}{2017}]{GolovinKrauseCorrected}
Golovin, D.\BBACOMMA\  \BBA\ Krause, A. \BBOP2017\BBCP.
\newblock \BBOQ Adaptive submodularity: {A} new approach to active learning and
  stochastic optimization (version 5)\BBCQ\
\newblock {\Bem CoRR}, {\Bem abs/1003.3967}.

\bibitem[\protect\BCAY{Hellerstein\ \BBA\ Kletenik}{Hellerstein\ \BBA\
  Kletenik}{2018}]{hellerstein:jair}
Hellerstein, L.\BBACOMMA\  \BBA\ Kletenik, D. \BBOP2018\BBCP.
\newblock \BBOQ Revisiting the approximation bound for stochastic submodular
  cover\BBCQ\
\newblock {\Bem J. Artif. Intell. Res.}, {\Bem 63}, 265--279.

\bibitem[\protect\BCAY{Hochbaum}{Hochbaum}{1982}]{hochbaumDualGreedy}
Hochbaum, D.~S. \BBOP1982\BBCP.
\newblock \BBOQ Approximation algorithms for the set covering and vertex cover
  problems\BBCQ\
\newblock {\Bem SIAM Journal on Computing}, {\Bem 11\/}(3), 555--556.

\bibitem[\protect\BCAY{Im}{Im}{2016}]{im:encyc16}
Im, S. \BBOP2016\BBCP.
\newblock \BBOQ Min-sum set cover and its generalizations\BBCQ\
\newblock In {\Bem Encyclopedia of Algorithms}, \BPGS\ 1331--1334. Springer.

\bibitem[\protect\BCAY{Im, Nagarajan,\ \BBA\ van~der Zwaan}{Im
  et~al.}{2016}]{imetal}
Im, S., Nagarajan, V., \BBA\ van~der Zwaan, R. \BBOP2016\BBCP.
\newblock \BBOQ Minimum latency submodular cover\BBCQ\
\newblock {\Bem {ACM} Trans. Algorithms}, {\Bem 13\/}(1), 13:1--13:28.

\bibitem[\protect\BCAY{Liu, Parthasarathy, Ranganathan,\ \BBA\ Yang}{Liu
  et~al.}{2008}]{Liu:2008:NAS:1376616.1376633}
Liu, Z., Parthasarathy, S., Ranganathan, A., \BBA\ Yang, H. \BBOP2008\BBCP.
\newblock \BBOQ Near-optimal algorithms for shared filter evaluation in data
  stream systems\BBCQ\
\newblock In {\Bem Proceedings of the ACM SIGMOD International Conference on
  Management of Data}, \BPGS\ 133--146.

\bibitem[\protect\BCAY{Nan\ \BBA\ Saligrama}{Nan\ \BBA\
  Saligrama}{2017}]{nanSaligrama}
Nan, F.\BBACOMMA\  \BBA\ Saligrama, V. \BBOP2017\BBCP.
\newblock \BBOQ Comments on the proof of adaptive stochastic set cover based on
  adaptive submodularity and its implications for the group identification
  problem in "{G}roup-based active query selection for rapid diagnosis in
  time-critical situations"\BBCQ\
\newblock {\Bem {IEEE} Trans. Information Theory}, {\Bem 63\/}(11), 7612--7614.

\bibitem[\protect\BCAY{Wan, Du, Pardalos,\ \BBA\ Wu}{Wan
  et~al.}{2010}]{WanDPW10}
Wan, P., Du, D., Pardalos, P.~M., \BBA\ Wu, W. \BBOP2010\BBCP.
\newblock \BBOQ Greedy approximations for minimum submodular cover with
  submodular cost\BBCQ\
\newblock {\Bem Comp. Opt. and Appl.}, {\Bem 45\/}(2), 463--474.

\bibitem[\protect\BCAY{Wolsey}{Wolsey}{1982}]{wolsey:submod}
Wolsey, L.~A. \BBOP1982\BBCP.
\newblock \BBOQ An analysis of the greedy algorithm for the submodular set
  covering problem\BBCQ\
\newblock {\Bem Combinatorica}, {\Bem 2\/}(4), 385--393.

\end{thebibliography}
\bibliographystyle{theapa}
\end{document}